\documentclass[longauth]{aa} 

\usepackage{color}

\newcommand{\kms}{km~s$^{-1}$}
\newcommand{\teff}{$T_{\mathrm{eff}}$}
\newcommand{\vsini}{$v\sin{i_\star}$}
\newcommand{\vsys}{$V_\mathrm{Sys}$}

\usepackage{graphicx}
\usepackage{txfonts}
\usepackage{hyperref}
\begin{document}

   \title{The GAPS programme at TNG\fnmsep\thanks{Based on observations made with the Italian Telescopio Nazionale Galileo (TNG) operated by the Fundaci\'{o}n Galileo Galilei (FGG) of the Istituto Nazionale di Astrofisica (INAF) at the Observatorio del Roque de los Muchachos (La Palma, Canary Islands, Spain).}}

   \subtitle{XXX. Atmospheric Rossiter-McLaughlin effect and atmospheric dynamics of \object{KELT-20b}}

   \author{M. Rainer
          \inst{1}
          \and F. Borsa\inst{2}
          \and L. Pino\inst{1,3}
          \and G. Frustagli\inst{2,4}
          \and M. Brogi\inst{5,6,7}
          \and K. Biazzo\inst{8}
          \and A.S. Bonomo\inst{6}
          \and I. Carleo\inst{9,10}
          \and R. Claudi\inst{10}
          \and R. Gratton\inst{10}
          \and A.F. Lanza\inst{11}
          \and A. Maggio\inst{12}
          \and J. Maldonado\inst{12}
          \and L. Mancini\inst{13,14,6}
          \and G. Micela\inst{12}
          \and G. Scandariato\inst{11}
          \and A. Sozzetti\inst{6}
          \and N. Buchschacher\inst{15}
          \and R. Cosentino\inst{17}
          \and E. Covino\inst{16}
          \and A. Ghedina\inst{17}
          \and M. Gonzalez\inst{17}
          \and G. Leto\inst{11}
          \and M. Lodi\inst{17}
          \and A.F. Martinez Fiorenzano\inst{17}
          \and E. Molinari\inst{18}
          \and M. Molinaro\inst{19}
          \and D. Nardiello\inst{20,10}
          \and E. Oliva\inst{1}
          \and I. Pagano\inst{11}
          \and M. Pedani\inst{17}
          \and G. Piotto\inst{21}
          \and E. Poretti\inst{17}
                    }

   \institute{INAF - Osservatorio Astrofisico di Arcetri, Largo Enrico Fermi 5, I-50125 Firenze, Italy\\ 
    \email{monica.rainer@inaf.it}
    \and INAF - Osservatorio Astronomico di Brera, Via E. Bianchi, 46, I-23807 Merate (LC), Italy 
    \and Anton Pannekoek Institute for Astronomy, University of Amsterdam Science Park 904 1098 XH Amsterdam, The Netherlands 
    \and Universit\`{a} degli Studi di Milano Bicocca, Piazza dell'Ateneo Nuovo, 1, I-20126 Milano, Italy 
    \and Department of Physics, University of Warwick, Coventry CV4 7AL, UK 
    \and INAF - Osservatorio Astrofisico di Torino, Via Osservatorio 20, I-10025 Pino Torinese (TO), Italy 
    \and Centre for Exoplanets and Habitability, University of Warwick, Gibbet Hill Road, Coventry CV4 7AL, UK 
    \and INAF - Osservatorio Astronomico di Roma, Via Frascati 33, I-00078 Monte Porzio Catone (Roma), Italy 
    \and Astronomy Department and Van Vleck Observatory, Wesleyan University, Middletown, CT 06459, USA 
    \and INAF - Osservatorio Astronomico di Padova, Vicolo dell'Osservatorio, 5, I-35122 Padova (PD), Italy 
    \and INAF - Osservatorio Astrofisico di Catania, Via S.Sofia 78, I-95123 Catania, Italy 
    \and INAF - Osservatorio Astronomico di Palermo, Piazza del Parlamento, 1, I-90134 Palermo, Italy 
    \and Department of Physics, University of Rome ``Tor Vergata'', Via della Ricerca Scientifica 1, I-00133, Rome, Italy 
    \and Max Planck Institute for Astronomy, K\"{o}nigstuhl 17, D-69117, Heidelberg, Germany 
    \and Department of Astronomy, University of Geneva, Chemin des Maillettes 51, CH-1290 Versoix, Suisse 
    \and INAF - Osservatorio Astronomico di Capodimonte, Salita Moiariello 16, I-80131 Napoli, Italy 
    \and INAF - Fundaci\'{o}n Galileo Galilei, Rambla Jos\'{e} Ana Fernandez P\'{e}rez 7, 38712 Bre{\~n}a Baja (TF), Spain 
    \and INAF - Osservatorio Astronomico di Cagliari, Via della Scienza 5, I-09047 Cuccuru Angius, Selargius (CA), Italy 
    \and INAF - Osservatorio Astronomico di Trieste, Via Giambattista Tiepolo, 11, I-34131 Trieste, Italy 
    \and Aix-Marseille Universit\'{e}, CNRS, CNES, LAM, Marseille, France 
    \and Dipartimento di Fisica e Astronomia Galileo Galilei -- Universit$\grave{\rm a}$ di Padova, Vicolo dell'Osservatorio 2, I-35122, Padova, Italy
             }

   \date{Received <date>; accepted <date>}

 
  \abstract
   {Transiting ultra-hot Jupiters are ideal candidates to study the exoplanet atmospheres and their dynamics, particularly by means of high-resolution, high signal-to-noise ratio spectra. One such object is KELT-20b, orbiting the fast rotating A2-type star \object{KELT-20}. Many atomic species have already been found in its atmosphere, with blueshifted signals that hints at the presence of a day-to-night side wind.}
   {We aimed to observe the atmospheric Rossiter-McLaughlin effect in the ultra-hot Jupiter KELT-20b, and to study any variation of the atmospheric signal during the transit. For this purpose, we analysed five nights of HARPS-N spectra covering five transits of KELT-20b.}
   {We computed the mean line profiles of the spectra with a least-squares deconvolution using a stellar mask obtained from the Vienna Atomic Line Database (\teff=10\,000 K, $\log g$=4.3), and then we extracted the stellar radial velocities by fitting them with a rotational broadening profile in order to obtain the radial velocity time-series. We used the mean line profile residuals tomography to analyse the planetary atmospheric signal and its variations.
   We also used the cross-correlation method to study an already known double-peak feature in the \texttt{FeI} planetary signal.}
   {We observed both the classical and the atmospheric Rossiter-McLaughlin effect in the radial velocity time-series. The latter gave us an estimate of the radius of the planetary atmosphere that correlates with the stellar mask used in our work ($R_{p+atmo}/R_p = 1.13 \pm 0.02$). We isolated the planetary atmospheric trace in the tomography, and we found radial velocity variations of the planetary atmospheric signal during transit with an overall blueshift of $\approx$ 10 \kms, along with small variations in the signal's depth and, less significant, in the full width at half maximum (FWHM).
   We also find a possible variation in the structure and position of \texttt{FeI} signal in different transits.}
   {We confirm the previously detected blueshift of the atmospheric signal during the transit. The FWHM variations of the atmospheric signal, if confirmed, may be caused by more turbulent condition at the beginning of the transit, or by a variable contribution of the elements present in the stellar mask to the overall planetary atmospheric signal, or by iron condensation. The \texttt{FeI} signal show hints of variability from one transit to the other.}

   \titlerunning{The GAPS programme at TNG XXX. Atmospheric RML and dynamics of KELT-20b}
   \maketitle
%

\section{Introduction}\label{sec:intro}
A very interesting category of exoplanets is represented by the transiting ultra-hot Jupiters (UHJs; \citealt{Bell2018}), which are highly irradiated Jupiter-size planets with day-side temperatures higher than 2200~K \citep{Parmentier2018}.
Transiting UHJs are ideal laboratories to study planetary atmospheres: their inflated atmospheres and high equilibrium temperatures ($T_\mathrm{eq}$)  result in strong signals and striking peculiar conditions for a planetary body. Their atmospheres are rich in atomic and molecular species: for example, \texttt{CrII}, \texttt{FeI}, \texttt{FeII}, \texttt{MgII}, \texttt{NaI}, \texttt{ScII}, \texttt{TiII}, and \texttt{YII} have been detected in KELT-9b, the hottest UHJ known so far ($T_\mathrm{eq} = 4050$~K), with additional evidence of the presence of \texttt{CaI}, \texttt{CrI}, \texttt{CoI}, and \texttt{SrII} \citep{Hoeijmakers2018, Hoeijmakers2019}.
Because of the presence of neutral and ionised iron in their atmospheres, UHJs can be used to study the atmospheric Rossiter-McLaughlin effect \citep{Borsa2019}: in fact the signal coming from their atmosphere correlates with the stellar mask used to compute the mean line profile of the spectra and recover the star's radial velocity (RV), resulting in an additional absorption in the mean line profile that causes an apparent RV variation similar to the classical Rossiter-McLaughlin effect (RML).
In addition to that, UHJs have usually very different atmospheric conditions (e.g., in temperature and chemical composition) between day and night side, that may result in day-to-night side wind \citep{Ehrenreich2020, Heng2015}: high resolution spectroscopy may be used to study the RV variations of the atmospheric signal in order to search for the presence of winds or any other kind of atmospheric turbulence.

KELT-20b \citep{Lund2017}, \textit{aka} MASCARA-2b \citep{Talens2018}, is a well known ultra-hot Jupiter orbiting a fast rotating A-type star. With a period of 3.47 days and a semi-major axis of 0.0542 au, KELT-20b is highly irradiated by its host star (A2, $T_\mathrm{eff} = 8980$~K, $m_V = 7.6$) and its atmosphere reaches $T_\mathrm{eq}$ = 2260~K (see Table~\ref{table:system} for more details on the system).

\begin{table*}

\caption{Physical and orbital parameters of the KELT-20 system (\textit{aka} MASCARA-2, \textit{aka} HD185603)}             
\label{table:system}      
\centering                          
\begin{tabular}{c c c}        
\hline\hline                 
Parameter & Symbol & Value \\    
\hline    
Stellar Parameters &  & \\
\hline
   Spectral type $^1$ &  & A2 \\
   $V$-band magnitude $^2$ & $m_V$  & 7.6 \\
   Effective temperature $^3$ & \teff  &  $8980^{+90}_{-130}$~K \\
   Projected rotation speed $^4$ & \vsini\  & $116.23\pm1.25$~\kms \\
   Linear limb darkening $^1$ & $u$  &  $0.532^{+0.011}_{-0.014}$ \\
   Surface gravity $^3$ & $\log g$  &  4.31 $\pm$ 0.02~cgs  \\
   Metallicity  $^3$ & [Fe/H] & -0.02 $\pm$ 0.07 dex\\ 
   Stellar mass $^3$ & $M_{\star}$  & $1.89^{+0.06}_{-0.05}~M_{\sun}$  \\ 
   Stellar radius $^3$ & $R_\star$  & $1.60 \pm 0.06~R_\sun$\\
   Rotation period $^4$ & $p_\star$ & 0.695 $\pm$ 0.027 days\\
\hline
Planetary parameters &  & \\
\hline
   Planet mass $^1$ & $M_p$ &  $<~3.51~M_{\mathrm{Jup}}$  \\ 
   Planet radius $^3$ & $R_p$  &  $1.83 \pm 0.07~R_{\mathrm{Jup}}$ \\
   Planet-to-star ratio $^1$ & $R_p/R_\star$  & $0.11440^{+0.00062}_{-0.00061}$\\
   Planet-to-star ratio $^3$ & $R_p/R_\star$  & $0.115 \pm 0.002$\\
   Equilibrium temperature $^3$ & $T_\mathrm{eq}$  & 2260 $\pm$ 50~K \\
   Surface gravity $^1$ & $\log g_\mathrm{p}$  & $<~3.46$~cgs \\
   Overall Fe volume mixing ratio (solar value) $^5$ & $\log \mathrm{VMR_{Fe}}$ & -4.27~cgs \\ 
\hline
Orbital parameters &  & \\
\hline
    Epoch $^3$ & $T_P$  & $2\,457\,909.5906^{+0.0003}_{-0.0002}$~BJD \\
    Period $^3$ & $P$  & $3.474119^{+0.000005}_{-0.000006}$~days \\
    Transit duration $^1$ & $T_\mathrm{dur}$  & $0.14882^{+0.00092}_{-0.00090}$~days \\
    Semi-major axis $^1$ & $a$ & $0.0542^{+0.0014}_{-0.0021}$~au\\
    Inclination $^1$ & $i$ & $86.15^{+0.28}_{-0.27}$~deg\\
    Eccentricity & $e$ & 0 (fixed)\\
    Projected obliquity $^3$ & $\lambda$  & 0.6 $\pm$ 4 deg\\
    Stellar RV amplitude $^6$ & $K_s$ & 322.51 m s$^{-1}$ \\
    Systemic velocity $^1$ & \vsys  & $-23.3 \pm 0.3$~\kms \\
    Systemic velocity $^3$  & \vsys & $-21.3 \pm 0.4$~\kms \\
    Systemic velocity $^7$ & \vsys  & $-22.06 \pm 0.35$~\kms \\
    Systemic velocity $^4$ & \vsys  & $-24.48 \pm 0.04$~\kms \\
\hline                                   
\end{tabular}
\tablebib{$^1$~\cite{Lund2017}; $^2$~\cite{Hog2000}; $^3$~\cite{Talens2018}; $^4$~this work; $^5$~\cite{Asplund2009}; $^6$~\cite{Casasayas2019}; $^7$~\cite{Nugroho2020}}

\end{table*}

Many atomic species such as \texttt{FeI}, \texttt{FeII}, \texttt{CaII}, \texttt{NaI}, \texttt{HI} have been detected in its atmosphere through transit spectroscopy, while there are only tentative detections of \texttt{MgI} and \texttt{CrII} \citep{Casasayas2018, Casasayas2019, Hoeijmakers2020, Nugroho2020, Stangret2020}.
There are also hints of the presence of a day-to-night side wind due to the presence of a blueshift of $-6.3 \pm 0.8$~\kms\ in the \texttt{FeI} signal, and $-2.8 \pm 0.8$~\kms\ in that due to \texttt{FeII} \citep{Stangret2020, Hoeijmakers2020, Nugroho2020}.

We observed KELT-20 in the framework of the Global Architecture of Planetary Systems (GAPS) project, which is an Italian project dedicated to the search and characterization of exoplanets (PI G. Micela; \citealt{Covino2013}). Particularly, one of GAPS' main lines of research  focuses on the study of exoplanets' atmospheres using both transmission and emission spectroscopy \citep{Borsa2019, Pino2020, Guilluy2020}.
Using both our data and public data of KELT-20 taken with the same instrument (HARPS-N), we studied both the classical \citep{Rossiter1924, McLaughlin1924} and atmospheric RML effects of KELT-20b from the RV time-series, along with the variations of the atmospheric trace during the planetary transits from the mean line profile tomography.

The dataset used in this work is described in Sec.~\ref{sec:data}, while the method used to obtain the stellar mean line profiles and the RV time-series is detailed in Sec.~\ref{sec:lsd}. Both the classical and atmospheric RML effects are shown in Sec.~\ref{sec:rml}. The variations of the atmospheric trace during the transit and the methods used to detect them from the mean line profile tomography are described in Sec.~\ref{sec:atmo_trace}. In Sec.~\ref{sec:ccf_fe} we describe the cross-correlation with a \texttt{FeI} model performed in order to compare our results with those found in the literature. Finally, our conclusions are in Sec.~\ref{sec:conclusion}.

\section{Data sample}\label{sec:data}
We analysed five transits of KELT-20b observed with the high-resolution echelle spectrograph HARPS-N \citep{Cosentino2012} installed at the Telescopio Nazionale Galileo (TNG) at the Roque de los Muchachos Observatory (La Palma, Spain).
HARPS-N is an optical spectrograph with resolving power $R=115\,000$ and wavelength coverage 383-693 nm. It is a twin of the HARPS spectrograph installed at the 3.6m telescope of the ESO-LaSilla Observatory, down to the Data Reduction Software (DRS) optimised for exoplanet search.

We observed two transits (2019 August 26 and 2019 September 02) in the framework of the GAPS programme, while the other three transits are public data retrieved from the HARPS-N archive (2017 August 16: PID CAT17A\_38 PI Rebolo; 2018 July 12, and 2018 July 19: PID CAT18A\_34, PI Casasayas-Barris).
The GAPS observations were taken in the GIARPS mode \citep{Claudi2016}, that allows the simultaneous use of both HARPS-N and GIANO-B \citep{Oliva2012, Origlia2014} spectrographs. GIANO-B is an high-resolution ($R = 50\,000$) near-infrared echelle spectrograph covering the wavelength range from 950 to 2450 nm. For this work, we used only the HARPS-N data

A summary of the acquired HARPS-N spectra in the five transit nights is shown in Table~\ref{table:data}. We rejected 14 spectra taken during night 2 because of their low S/N. All five transits are complete, and out-of-transit spectra were taken both before and after the transit in each night.
\begin{table}
\caption{Data summary. Throughout the paper the transits will be identified by their number (first column).}
\label{table:data}      
\centering                          
\begin{tabular}{c c c c c}        
\hline\hline                 
$\#$ & Night & $\#$ of spectra & $T_\mathrm{exp}$ & Mean S/N \\    
\hline                        
   1 & 2017 August 16 & 90 & 200s & 61 \\   
   2 & 2018 July 12 & 116 & 200s & 93 \\
   3 & 2018 July 19 & 78 & 300s & 105 \\
   4 & 2019 August 26 & 30 & 600s & 164 \\
   5 & 2019 September 02 & 29 & 600s & 176 \\ 
\hline                                   
\end{tabular}
\end{table}
We worked on spectra reduced by the HARPS-N DRS \citep{Cosentino2014}, as such the barycentric correction was already applied.


\section{Mean line profiles}\label{sec:lsd}
While the HARPS-N DRS is a very powerful tool, it is not optimised for hot stars such as KELT-20: the resulting cross-correlation functions (CCFs) are obtained by using a stellar mask designed to work with a G2-type star, which is the hottest stellar mask available in the DRS mask library.

\begin{figure}
    \centering
    \includegraphics[trim=0 0 0 38, clip, width=\columnwidth]{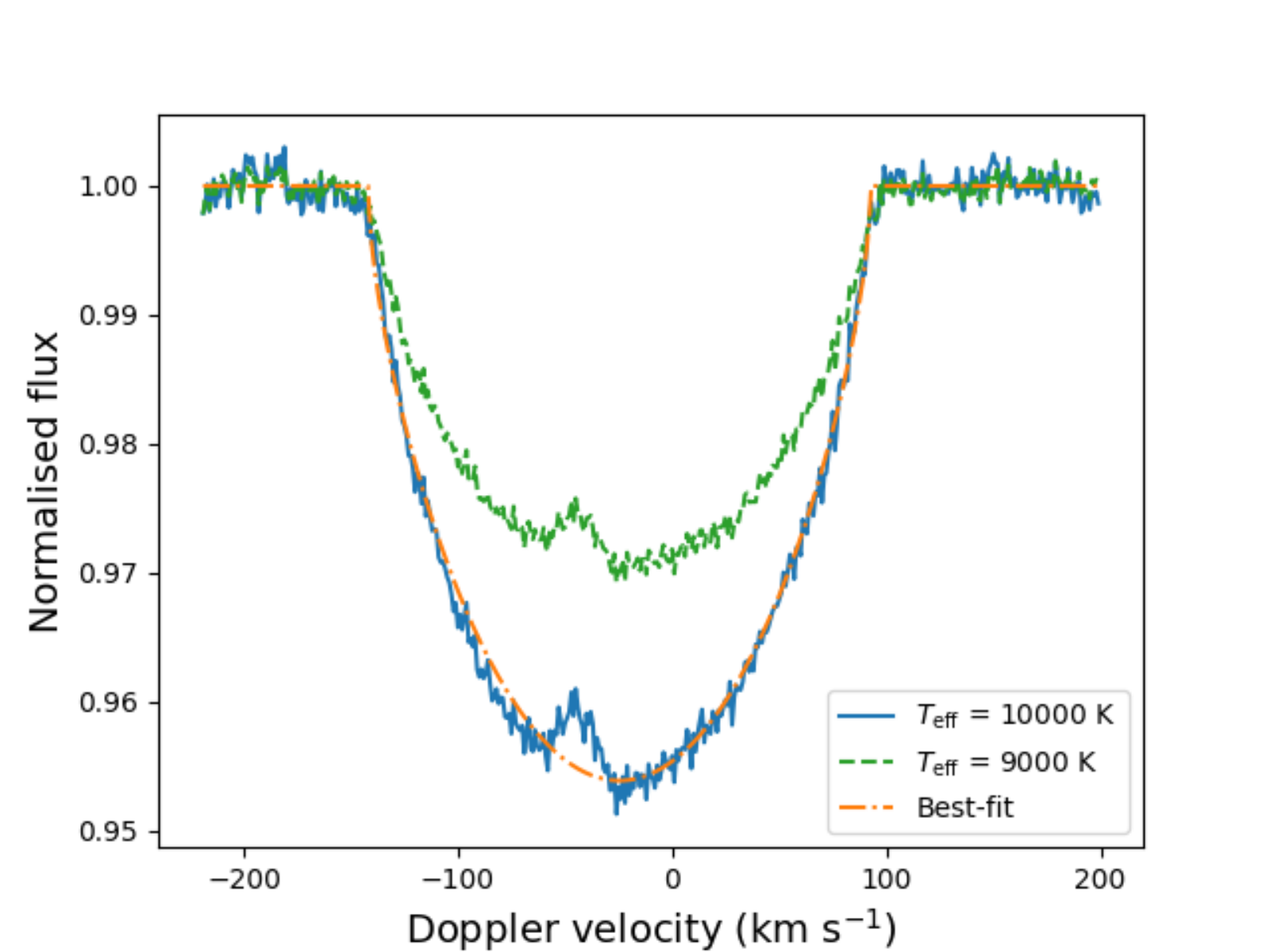}
    \caption{Mean line profile of a single KELT-20 spectrum obtained with the LSD software and \teff=10\,000~K (solid blue line)  along with the rotational broadening fit (dashed-dotted orange line). For comparison, the mean line profile of the same spectrum with the \teff=9000~K is shown (dashed green line). The planet's Doppler shadow is clearly visible as a bump in both lines.}
    \label{fig:fitLSD}
\end{figure}

We decided then to compute the mean line profile using the Least-Squares Deconvolution software (LSD, \cite{Donati1997}) with a stellar mask obtained from the VALD3 database\footnote{\url{http://vald.astro.uu.se}} \citep{Piskunov1995, Ryabchikova2015}. We downloaded stellar masks for \teff=9000~K and \teff=10\,000~K, both with $\log g$=4.31, solar metallicity, and micro-turbulence $\nu$=2~km~s$^{-1}$ and wavelength range 3900-7000~{\AA}.
While our results are in good agreement using both masks, we show here only those obtained with \teff=10\,000~K, where both the stellar and the planetary signals are stronger (see Fig.~\ref{fig:fitLSD}, where the \teff=10\,000~K profile is almost twice as deep as the \teff=9000~K one). This may hint at an higher $T_\mathrm{eff}$ for KELT-20 then previously found.

To apply the LSD, we first normalised all our spectra using a self-developed automated procedure \citep{Rainer2016}. 
Then, in order to avoid most of the telluric lines contamination and the heavy contribution of the stellar Balmer lines (extremely strong as expected from KELT-20 spectral type, and with a different shape), we cut them, and we kept only the wavelength ranges 4415-4805, 4915-5870, 6050-6265, and 6335-6450 {\AA}. We run the LSD software on each individual spectrum to obtain the mean line profiles.

We fitted all the mean line profiles with a rotational broadening function (see Fig.~\ref{fig:fitLSD}), using the formula in Eq.~\ref{eq:gray} \citep{Gray2008}:
\begin{equation}
    f(x) = 1-a \frac{2(1-u) \sqrt{1-\left( \frac{x-x_0}{x_l}\right) ^{2}} + 0.5\pi u \left[1- \left(\frac{x-x_0}{x_l}\right)^{2}\right]} {{\pi}x_l \left(1-\frac{u}{3} \right)},
    \label{eq:gray}
\end{equation}
where $a$ is the depth of the profile, $x_0$ the centre (i.e., the RV value), $x_l$ the \vsini\ of the star, $u$ the linear limb darkening (LD) coefficient (listed in Table~\ref{table:system}).
We thus recovered both the RVs and the projected rotational velocities (\vsini) for all observed spectra.
We found $v\sin{i_\star} = 116.7\pm0.7$~\kms\ by averaging the \vsini\ of all the out-of-transit spectra. We did not use here the in-transit ones in order to avoid the Doppler shadow affecting our result.

We also computed the \vsini\ using the Fourier transform method \citep{Smith1976, Dravins1990}: because of the fast rotation of KELT-20 we could use the first three zero positions of the Fourier transform of all our mean line profiles to derive the projected rotational velocity. Using only the out-of-transit spectra, we found an average value of $v\sin{i_\star} = 116.23\pm1.25$~\kms, which aligns well with that obtained by the profile fitting. We note here that, in case of such a fast rotating star, this method is independent from other broadening effects as for example macro-turbulence, and it only depends on the LD coefficient: for this reason, we report this value in Table~\ref{table:system}, even if the error is larger than that obtained with the profile fitting.
The average value of the ratio of the first two zero positions results in $q_2/q_1 = 1.805 \pm 0.036$, which is compatible with a rigid rotation ($1.72 < q_2/q_1 < 1.83$, \cite{Reiners2002}).

We computed the stellar rotational period as $p_\star = 2\pi R_\star / v_\mathrm{eq}$, using $R_\star$, \vsini, and inclination $i$ from Table~\ref{table:system}. We considered the orbit inclination $i$ equal to the stellar inclination $i_\star$, seeing as the projected obliquity $\lambda$ is compatible with a zero value. We found a stellar rotational period of 0.695 $\pm$ 0.027 days, which is almost exactly one fifth of the planetary period: this may suggests a resonance between the stellar and planetary rotation. 

\begin{figure}
    \centering
    \includegraphics[trim=0 0 0 38, clip, width=\columnwidth]{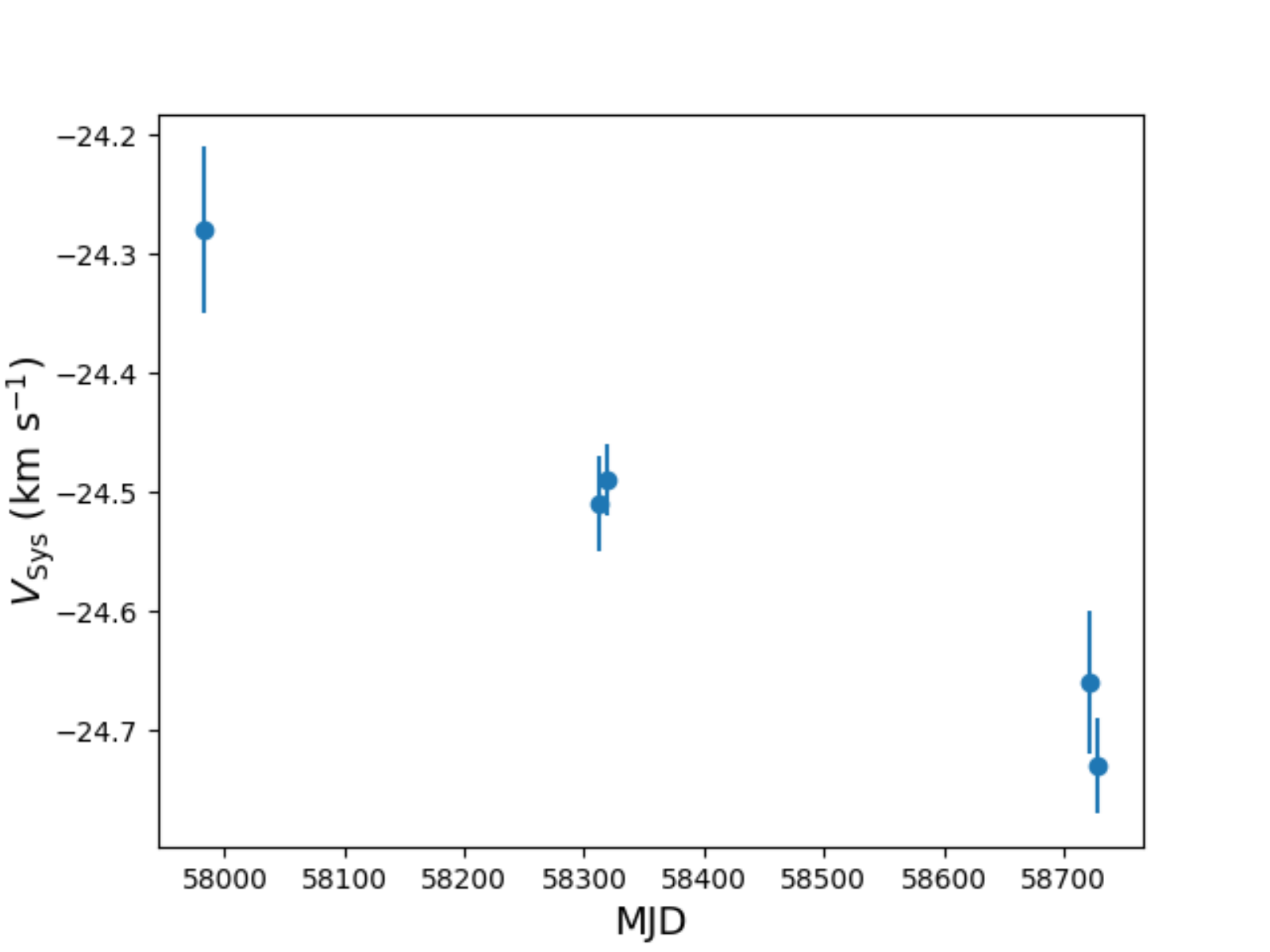}
    \caption{Systemic velocity of the five nights: a small downwards trend is clearly visible.}
    \label{fig:vsys}
\end{figure}

We performed a linear fit on all the out-of-transit RVs to recover the systemic velocity, and we found $V_\mathrm{Sys} = -24.48 \pm 0.04$ \kms. This value is slightly lower than those found in the literature (see Table~\ref{table:system}), but the determination of the systemic velocity may vary depending on the instrument and method used to estimate it.
We also computed \vsys\ independently on the five nights, and we noted a small, but significant downwards trend that may be worth keeping in mind in further studies (see Fig.~\ref{fig:vsys}).

\begin{figure}
    \centering
    \includegraphics[trim=0 0 0 38, clip,  width=\columnwidth]{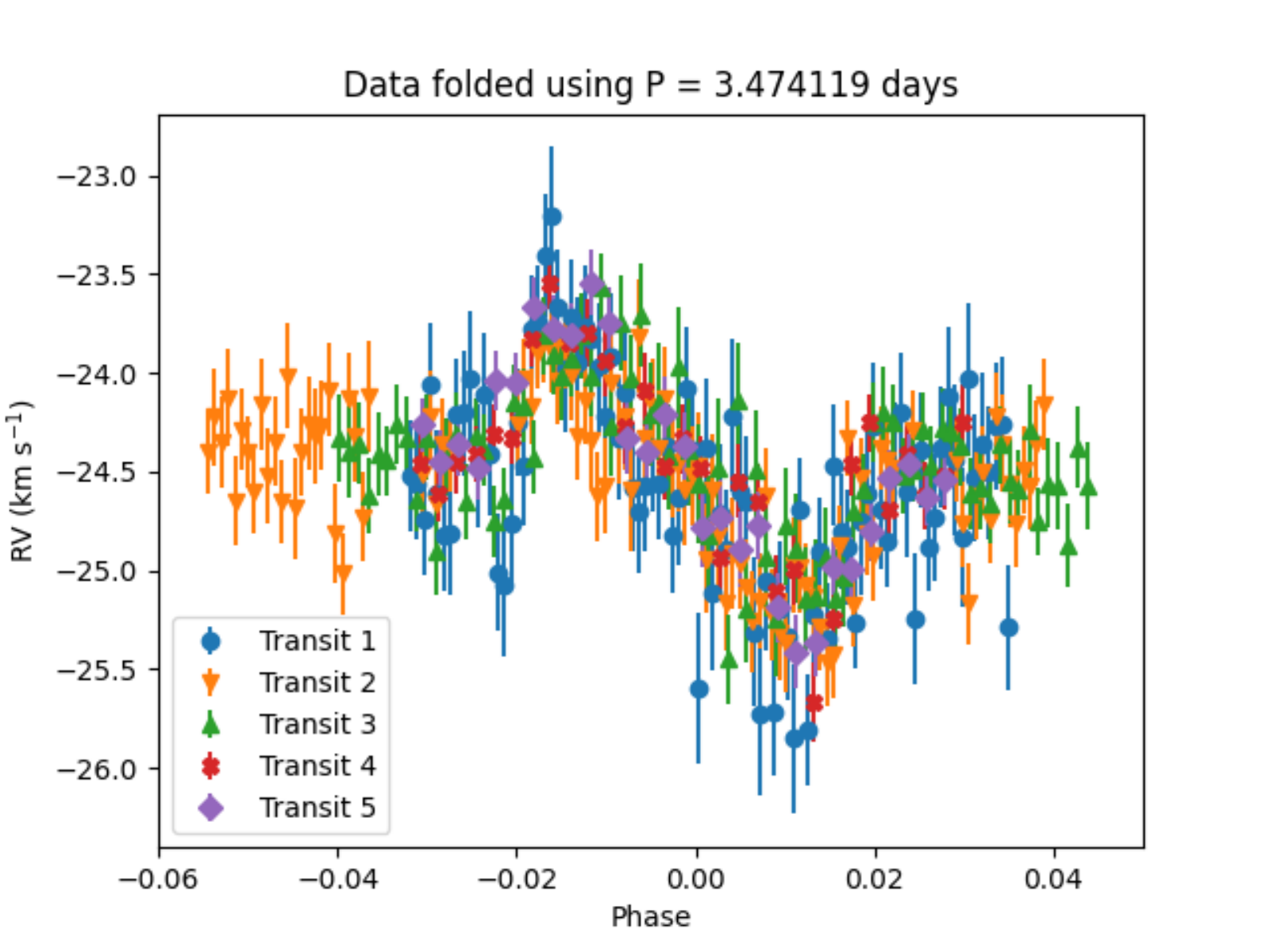}
    \caption{RVs data phase-folded using the KELT-20b known orbital period ($P=3.474119$ days). The RV values were shifted by the difference between the individual \vsys\ of each night and the average \vsys\ value to account for the trend found in our data.}
    \label{fig:RMLall}
\end{figure}

In Fig.~\ref{fig:RMLall} we show the RVs corrected for the different \vsys\ (so that the five nights aligns on the average \vsys\ value), and phase-folded using the known orbital value of KELT-20b ($P=3.474119$ days, see Table~\ref{table:system}). The RML effect is clearly visible.

\section{Classical and atmospheric RML effects}\label{sec:rml}
The RML effect is visible in all nights of observations (see Fig.~\ref{fig:RMLall}).
We averaged the phase-folded RVs data of all transits using a 0.002 phase bin and we compared them with a theoretical model obtained using the already known system parameters of Table~\ref{table:system}, using our value for the systemic velocity.
The RML model was computed with the \texttt{Rmcl} model class of the PyAstronomy\footnote{\url{https://github.com/sczesla/PyAstronomy}} package \citep{pya} of Python\footnote{\url{http://www.python.org}} \citep{van1995python}, which implements the analytical model RV curves for the RML effect given by \cite{Ohta2005}.

\begin{figure}
    \centering
    \includegraphics[trim=0 0 0 38, clip, width=\columnwidth]{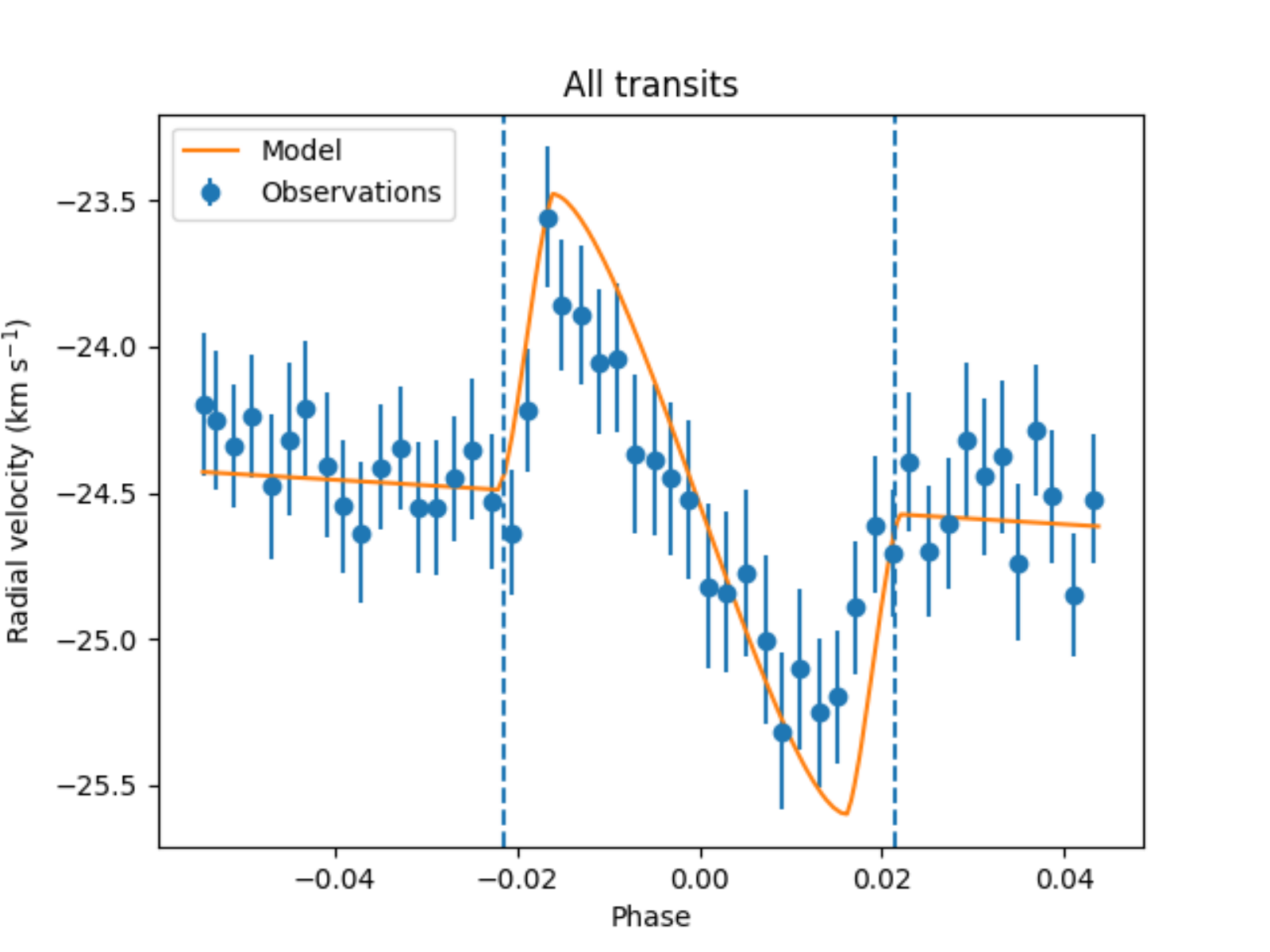}
    \caption{Comparison between the observed RVs averaged with 0.002 phase bin (blue points) and the RML RV model (orange line). The model seems to overestimate the amplitude of the RML effect, while actually the atmospheric RML effect is lowering the amplitude of the signal. The vertical dashed lines show the transit's ingress and egress.}
    \label{fig:RMLmodel}
\end{figure}

The comparison between the data and the theoretical model is shown in Fig.~\ref{fig:RMLmodel}: the model seems to overestimate the amplitude of the RML effect, but we know from a previous study on KELT-9b \citep{Borsa2019} that the atmospheric RML effect may combine with the classical RML effect and the resulting RVs carry both signals.

The atmospheric RML effect is akin to the classical one: an apparent stellar RV variation due to the deformation of the stellar lines. While in the classical RML effect the deformation is caused by the occultation of part of the stellar disk by the transiting planet, in the atmospheric RML effect we have an additional absorption signal due to the planetary atmospheric spectrum correlating with the mask used to compute the CCF or the LSD mean line profile.
This happens only in the case of extremely hot planetary atmospheres, that show a chemical composition similar to that of late-type stars (in particular due to the presence of neutral or ionised iron), and as such their atmospheric spectrum correlates with the same mask used for the host stars (e.g., in this case the stellar mask contains most of the elements found in the atmosphere of KELT-20b, with more than half of the lines being either \texttt{FeI} or \texttt{FeII}).

\begin{figure}
    \centering
    \includegraphics[trim=0 0 0 38, clip, width=\columnwidth]{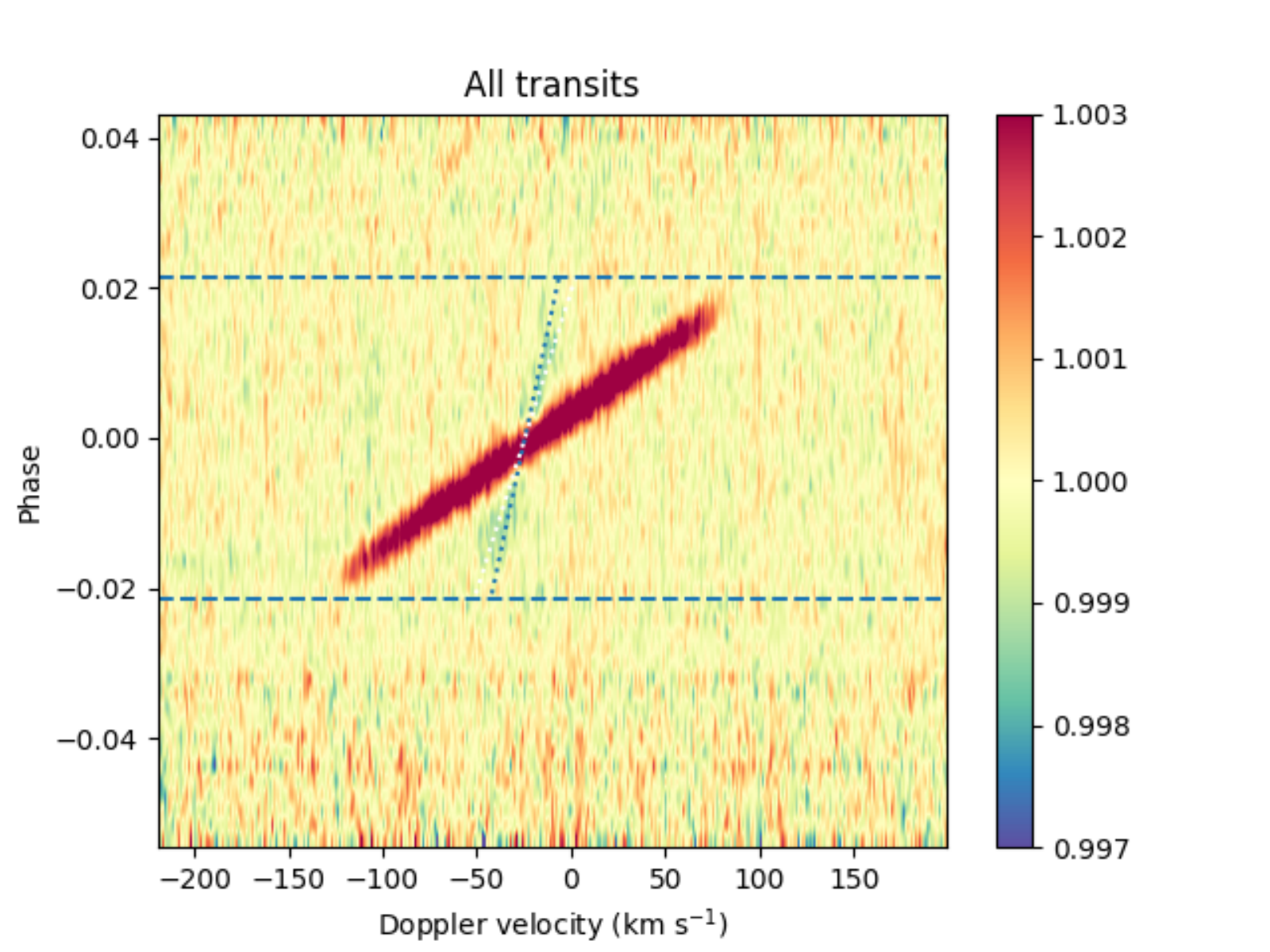}
    \caption{Mean line profile tomography: the average stellar line has been removed, but not the systemic velocity. In the residuals both the Doppler shadow (red excess) and the atmospheric trace (blue absorption, evidenced by the blue dotted line) are visible.}
    \label{fig:tomography}
\end{figure}

Looking at the line profile residuals tomography (Fig.~\ref{fig:tomography}, see Section~\ref{sec:atmo_trace} for details), not only the Doppler shadow is visible (red hues), but also the planetary atmospheric trace (blue hues). The latter shifts by the change in planet’s orbital RV during transit, confirming that the planet's atmospheric spectrum is correlating with the stellar mask, and thus it shows up in the line residuals as an additional absorption line situated at the planet RV.
The Doppler shadow and the atmospheric trace are aligned in such a way that the atmospheric trace is expected to affect the RVs derived from the mean line profiles in the opposite way than the Doppler shadow, so that the net result would be a smaller amplitude of the RML effect, as it is actually seen in Fig.~\ref{fig:RMLmodel}.

\begin{figure}
    \centering
    \includegraphics[trim=0 0 0 38, clip, width=\columnwidth]{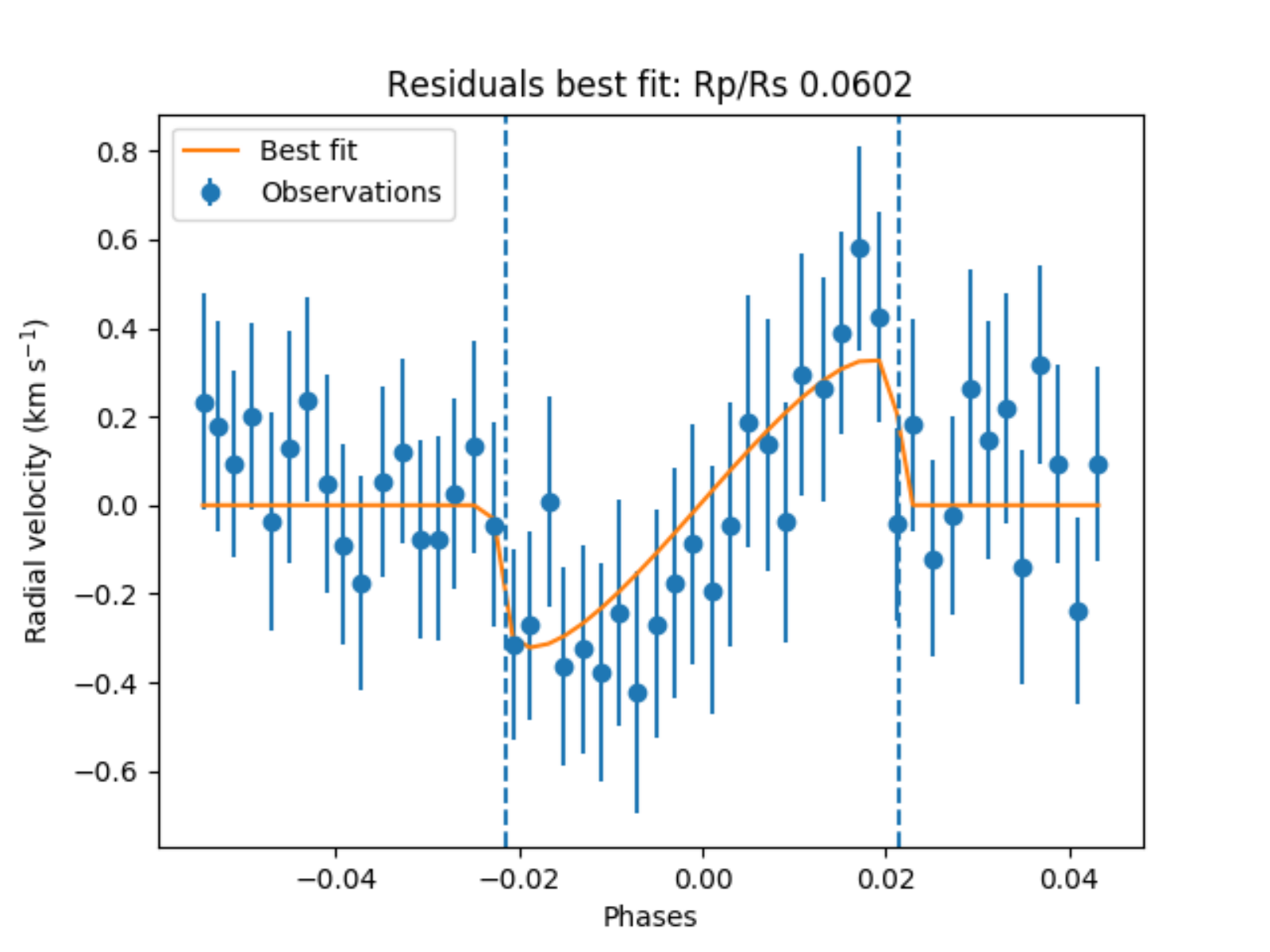}
    \caption{RVs residuals show the atmospheric RML effect. The overplotted orange line shows the RML model best fit.}
    \label{fig:atmoRML}
\end{figure}

We then subtracted the RML theoretical model from our data: the resulting RV residuals show the atmospheric RML effect. It goes in the opposite direction from the classical RML effect because it modifies the line profile as an additional absorption instead of a bump.
We fit the RV residuals (see Fig.~\ref{fig:atmoRML}) using the \texttt{Rmcl} fit class of the same PyAstronomy package used before.
We interpret the resulting value $R_p/R_\star = 0.060 \pm 0.002$ given by the fit as $R_{\mathrm{atmo}}/R_\star$, where $R_{\mathrm{atmo}}$ represents the extension of the atmosphere that correlates with the stellar mask that we used, if the atmosphere were shaped as a disk. The error on this value was estimated using the \texttt{pymc}\footnote{\url{https://github.com/pymc-devs/pymc}} package of Python.
Comparing this result with the planet's radius $R_p/R_\star = 0.115 \pm 0.002$, the atmospheric area is $\sim 27 \%$ of the whole planetary photometric area. If we hypothesize the most simple scenario of a spherical atmosphere, we can then derive $R_{p+atmo} = 1.13 \pm 0.02~R_p$, only for the portion of the planetary atmosphere whose spectrum correlates with the stellar mask. This value is in good agreement with the results from \cite{Casasayas2019}, who found $R_{p+atmo} = 1.11 \pm 0.03~R_p$ for \texttt{FeII}. The slightly larger value found here may be due to the presence of few lines of other elements, such as \texttt{CaII}, for which \cite{Casasayas2019} found $R_{p+atmo} = 1.19 \pm 0.03~R_p$.

We note here that adjusting the classical RML theoretical model by varying the system parameters values inside their error ranges alters only slightly our atmospheric RML result, which remains in agreement with the $R_{atmo}/R_\star = 0.060 \pm 0.002$ value within the $1\sigma$ uncertainty.


\section{Atmospheric trace}\label{sec:atmo_trace}
We studied the planetary atmospheric trace following the same strategy used with the RVs in Sec.~\ref{sec:rml}, and applying it to the mean line profiles: we combined all five transits, in order to increase the strength of the atmospheric signal and to average out spurious variations due to instrumental or telluric effects.

We averaged all our out-of-transit mean line profiles to obtain a purely stellar mean line profile. Because KELT-20 does not show any significant stellar variations, either due to activity or pulsations, we could then remove the stellar component from our data simply by dividing each mean line profile (both in and out of transit) by the average out-of-transit stellar mean line profile. We then normalised the residuals by dividing them using two different linear fits, one for the points outside the stellar line limits, the other for the points inside the stellar line limits. The latter fit was done avoiding the regions where the Doppler shadow or the atmospheric trace are present. The resulting residuals are shown in Fig.~\ref{fig:tomography}; to enhance the signal's visibility they are binned with a 0.002 phase bin and a 1~\kms\ RV bin.

To isolate and investigate possible variations of the atmospheric trace during transit, we had to remove the Doppler shadow. To be sure that the removal process did not influence our analysis, we proceeded in two different ways:
\begin{enumerate}
    \item[a)] by following the method of \cite{Hoeijmakers2019}: we selected the residuals where the Doppler shadow signal is far from the atmospheric trace and we fitted it with a Gaussian. Then we fitted the Gaussian parameters with a 2$^{nd}$ order polynomial, so that the fit parameters of the Doppler shadow vary smoothly during the transit. We then removed the Doppler shadow Gaussian model from all the residuals.
    
    We note here that we obtained a better removals by first shifting all our data in the reference frame of the Doppler shadow, probably because of the geometry of the KELT-20 system.
    We did this using the estimated Doppler shadow RV obtained from Eq.~\ref{eq:cegla} \citep{Cegla2016}:
    \begin{equation}
    \label{eq:cegla}
    rv = v\sin{i_\star} \left( x_p \cos{\lambda} - y_p \sin{\lambda}\right),
    \end{equation}
    where:
    \begin{equation*}
    x_p = a_{R_\star} \sin{2\pi \phi},
    \end{equation*}
    \begin{equation*}
    y_p = -a_{R_\star} \cos{2\pi \phi} \cos{i},
    \end{equation*}
    with $\lambda$ the projected obliquity in radians,  $a_{R_\star}$ the semi-major axis in units of stellar radius, $\phi$ the orbital phase and $i$ the orbital inclination in radians. After this, our Doppler shadow signal was vertically aligned, and we proceeded with the removal as described above;
    \item[b)] by adopting and adjusting the method from \cite{Cabot2020}. The original method was applied to the UHJ WASP-121b, which orbits in a near-polar orbit around its host star ($\lambda = 257.8^{+5.3}_{-5.5}$ deg; \citealt{Delrez2016}). Because of this, its Doppler shadow in the stellar reference frame is almost completely vertically aligned at the center of the stellar line profile. \cite{Cabot2020} removed it by fitting a $3^\mathrm{rd}$ degree polynomial on each column of the tomography where the Doppler shadow fell. 
    
    Because of the different geometry of the KELT-20 system, we had to modify this approach to suit our data. First of all, we shifted the data in the reference frame of the Doppler shadow as in the original work of \cite{Cabot2020}. We could then fit the columns where the Doppler shadow signal fell, and finally we divided each column by its fit. We used a $5^\mathrm{th}$ degree polynomial, instead of the original $3^\mathrm{rd}$ degree one, because it performed a better removal of the Doppler shadow.
\end{enumerate}

We obtained thus two datasets: dataset A, where the Doppler shadow was removed by Gaussian fitting, and dataset B, where the Doppler shadow was removed adapting the \cite{Cabot2020} method. Because the results we obtained with the two datasets are in good agreement, we show here only the work done on dataset A, while the results from dataset B are presented in appendix~\ref{app:cabot}.

\begin{figure}
    \centering
    \includegraphics[trim=0 0 0 38, clip, width=\columnwidth]{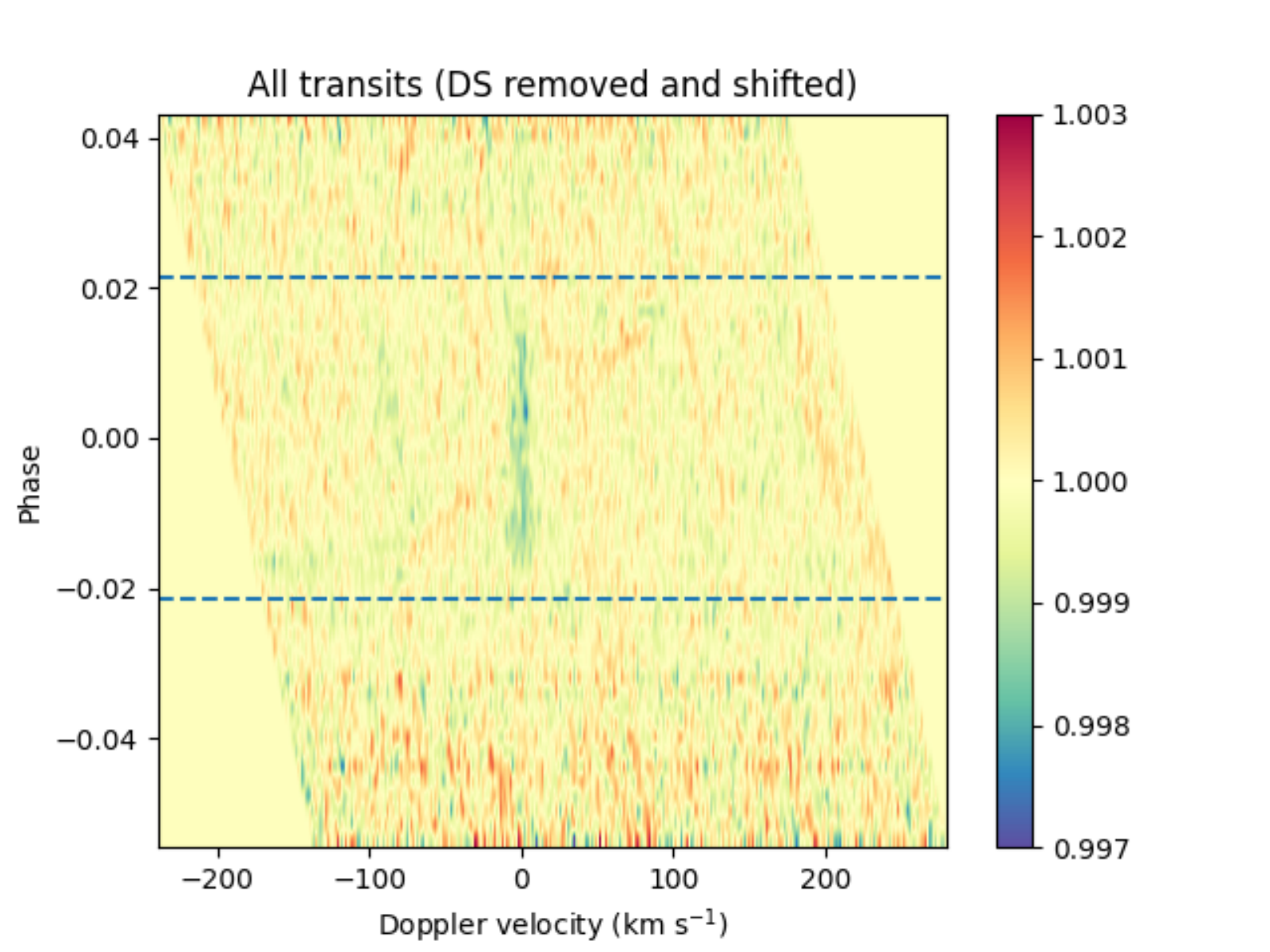}
    \caption{Line profile residuals after the Doppler shadow's removal, and after being shifted in the planet reference frame. The atmospheric trace is clearly visible and centered around 0~\kms.}
    \label{fig:atmo_trace}
\end{figure}

Once removed the Doppler shadow signal, we shifted the dataset in the planet reference frame: we shifted each spectrum by the combination of \vsys\ and the planet theoretical orbital RV (the barycentric correction was already applied by the HARPS-N DRS), in order to align the atmospheric signal in a vertical position around 0~\kms\ and to better study its variations (see Fig.~\ref{fig:atmo_trace}).
The planet orbital RVs were computed for each epoch with the \texttt{KeplerEllipse} class of the PyAstronomy package using the orbital values of semi-major axis $a$, period $P$, eccentricity $e$, and inclination $i$ from Table~\ref{table:system}. This yields a $K_p$ of $169 \pm 6$~\kms, compatible with the values found in literature \citep{Casasayas2019,Nugroho2020,Stangret2020}.

We then considered each mean line profile to map the velocity variations during transit. Because the atmospheric signal is not very strong, we applied a Savitzky-Golay filter \citep{Savitzky1964} to each profile in order to smooth out the noise and increase the signal's visibility. The Savitzky-Golay filter works by computing a least-squares low-degree polynomial fit ($3^\mathrm{rd}$ degree in our case) in a moving window on the data to estimate the value of the central point of each window, and it is able to smooth the data without greatly distorting the signal. While it was originally created for spectroscopic chemistry, the Savitzky-Golay filter has been successfully applied to several kind of spectroscopic astronomical data \citep[e.g.,][]{Deetjen2000,Dimitriadis2019,Fleig2008}. We applied the Savitzky-Golay filter by using the \texttt{savgol\_filter} function of SciPy\footnote{\url{https://www.scipy.org}} with an optimal window width of 15 pixel. We obtained the window value by applying the method proposed by \cite{Sadeghi2018} to our data. In appendix~\ref{app:savgol} we show the results of the atmospheric analysis on the unfiltered dataset A, to highlight the improvements obtained using the filter: the overall behaviour of the signal is clearly the same, but the uncertainties on the estimated values are much lower using the filtered data.

\begin{figure*}
    \centering
    \includegraphics[trim=0 0 0 38, clip, width=\textwidth]{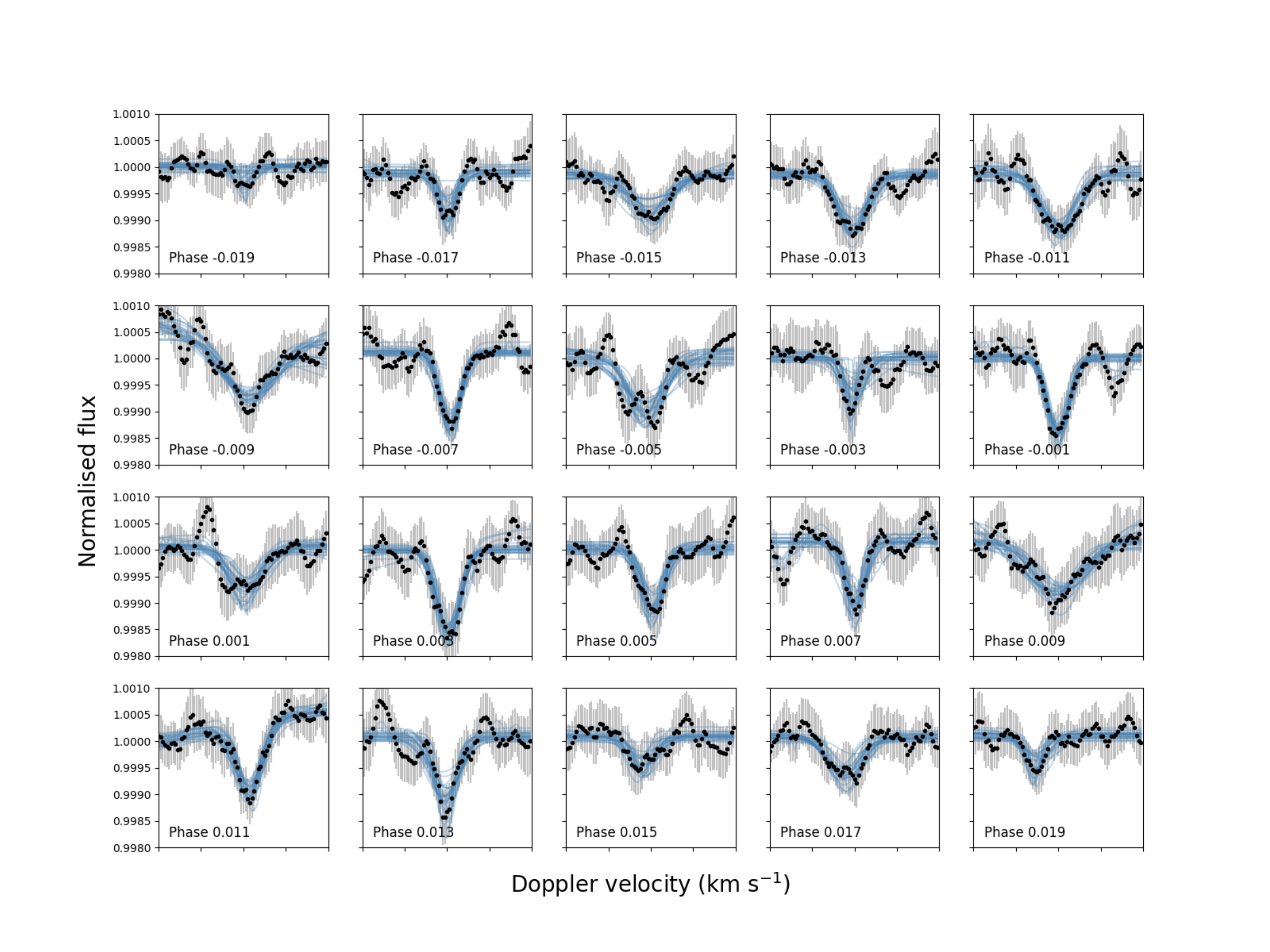}
    \caption{The atmospheric signal found in the line profile residuals and the relative MCMC posteriors.
    All the graphics have the same abscissa (RVs from -40 to 40 \kms) and ordinate (normalised flux from 0.998 to 1.001) to better follow the evolution of the signal. The graphics go from phase -0.019 (upper left panel) to phase 0.019 (lower right panel) with a 0.002 phase step.}
    \label{fig:fit_atmo_trace}
\end{figure*}

After applying the Savitzky-Golay filter, we fitted the atmospheric signal with a Markov-Chain MonteCarlo (MCMC) sampling and a correlated noise model using Gaussian processes with a Mat\'{e}rn-3/2 covariance kernel. In order to do this, we used the Python packages \texttt{emcee}\footnote{\url{https://emcee.readthedocs.io/en/stable/}} \citep{Foreman-Mackey2013} and \texttt{george}\footnote{\url{https://george.readthedocs.io/en/latest/}} for the noise model. The priors on FWHM, RV, and depth were estimated as an average of the Gaussian fits on the different datasets, with broader widths than the expected uncertainties: FWHM=$20\pm20$ \kms, RV=$-5\pm20$ \kms, the initial depth is computed on each dataset as depth=minimum-maximum, and it ranges from $3\times$depth to 0.1 (to account for normalisation problems).
In Fig.~\ref{fig:fit_atmo_trace} we show 24 random posteriors for each of the 20 mean line profile residuals that we have during the transit.

\begin{figure}
    \centering
    \includegraphics[trim=0 0 0 38, clip, width=\columnwidth]{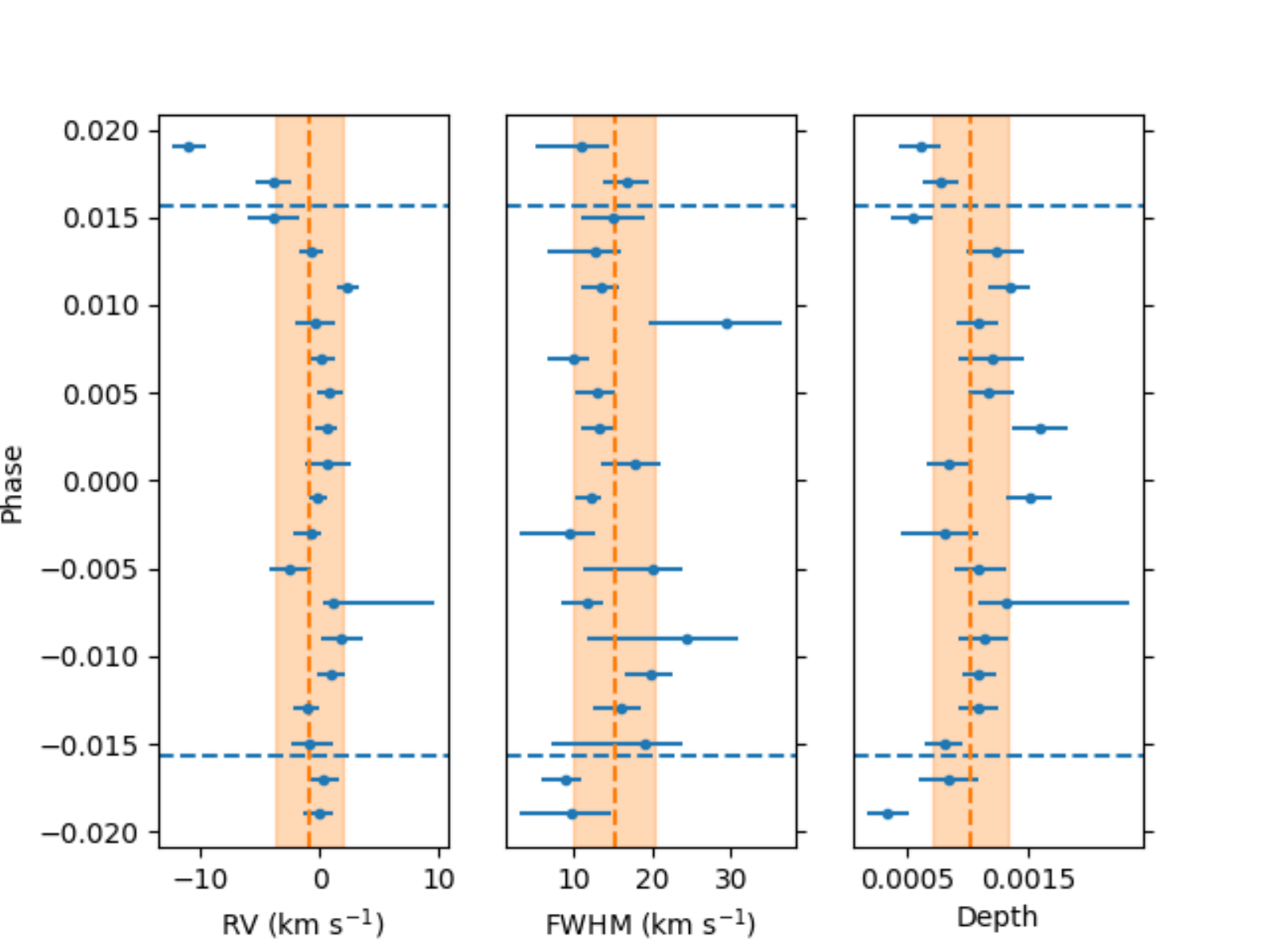}
    \caption{RVs, FWHM and depth of the atmospheric signal shown in Fig.~\ref{fig:fit_atmo_trace} during transit: the vertical dashed orange lines and shaded areas show the mean values and relative standard deviation for each of the three quantities. All the data are comprised between $t_0$ (start of ingress) and $t_2$ (end of egress), while the horizontal dashed blue lines indicate $t_2$ (end of ingress) and $t_3$ (start of egress).}
    \label{fig:atmo_results}
\end{figure}

The fit results are shown in Fig.~\ref{fig:atmo_results}: the 0.50, 0.16, and 0.84 quantiles of the posteriors distribution are used as the best values and the upper and lower 1$\sigma$ uncertainties respectively.
The vertical dashed orange lines and shaded areas show the mean values of RVs, FWHM and depth and the 1$\sigma$ regions.
The RV signal remain stable for most of the transit, and then it blueshifts during egress. Aside from one outlier, the FWHM is more stable: even if it seems to be lower during ingress there is no statistically evident variation from the mean. The depth of the signal increases from the beginning to the center of the transit, and then it decreases in a roughly symmetric way, with some points differing more than 1$\sigma$ from the mean during ingress and egress (a less deep atmospheric signal), and in the middle of the transit (a more deep signal). In all cases, the variations (if any) are mainly found either during ingress, or egress, or both.

\begin{figure}
    \centering
    \includegraphics[width=\columnwidth]{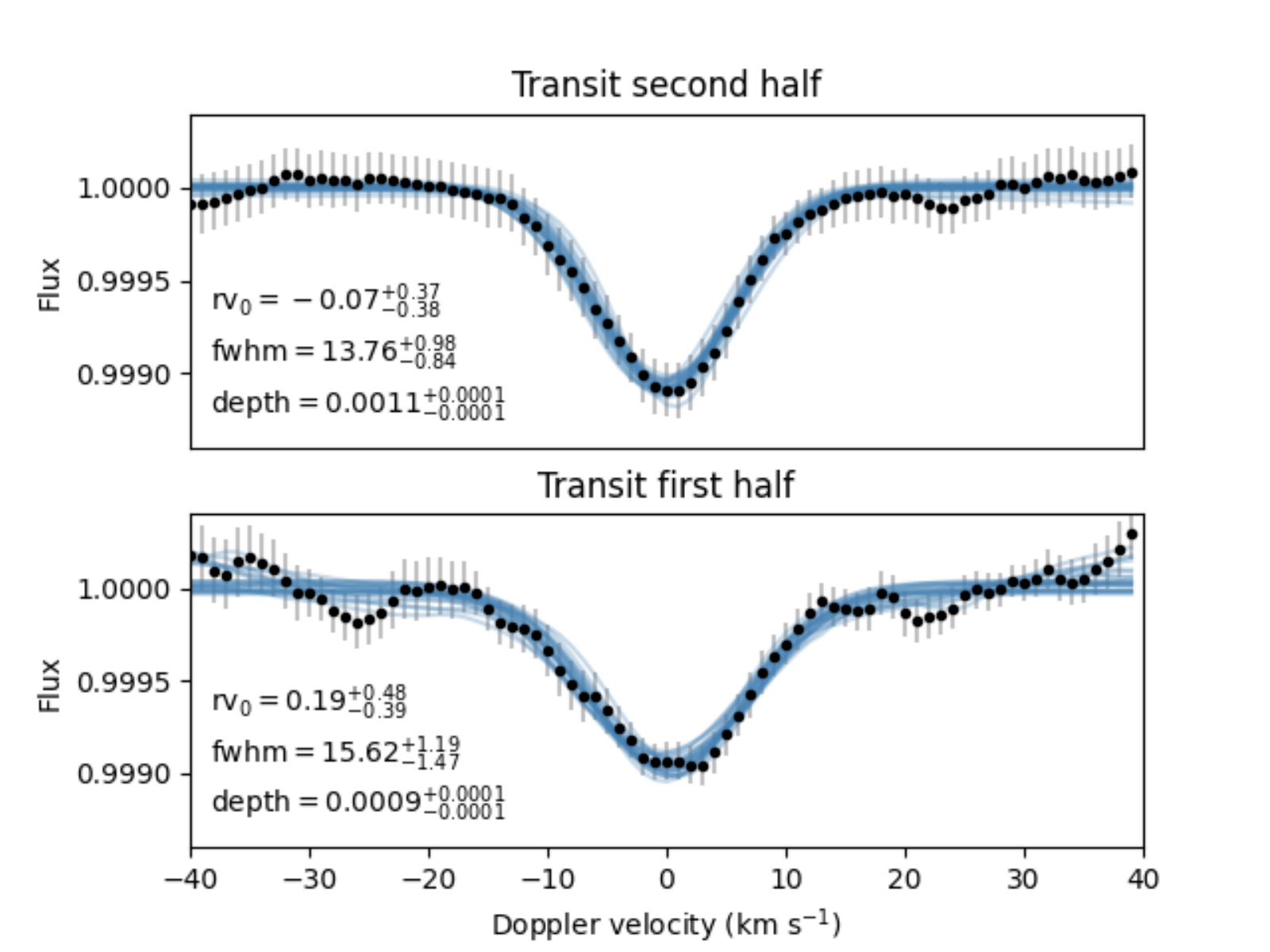}
    \caption{Averaged atmospheric signal found in the line profile residuals in the first (phase [-0.02:0.0]) and second part (phase [0.0:0.02]) of the combined transits with relative fit.}
    \label{fig:atmo_two_halves}
\end{figure}

Averaging the residuals in the first and the second half of the transit (phases [-0.02:0.0] and [0.0:0.02], see Fig.~\ref{fig:atmo_two_halves}) shows a larger FWHM value in the first half of the transit, but the result is not statistically significant. The depth variation is cancelled out within 1$\sigma$, and also the RV variation in the averaged signals disappears within 1$\sigma$. The latter is due to the fact that the major RV variation is caused only by the last three points (and the last one in particular), where the atmospheric signal is smaller (see last panels of Fig.~\ref{fig:fit_atmo_trace}) and as such their contribution to the average is lower.

\begin{figure*}
    \centering
    \includegraphics[width=0.3\textwidth]{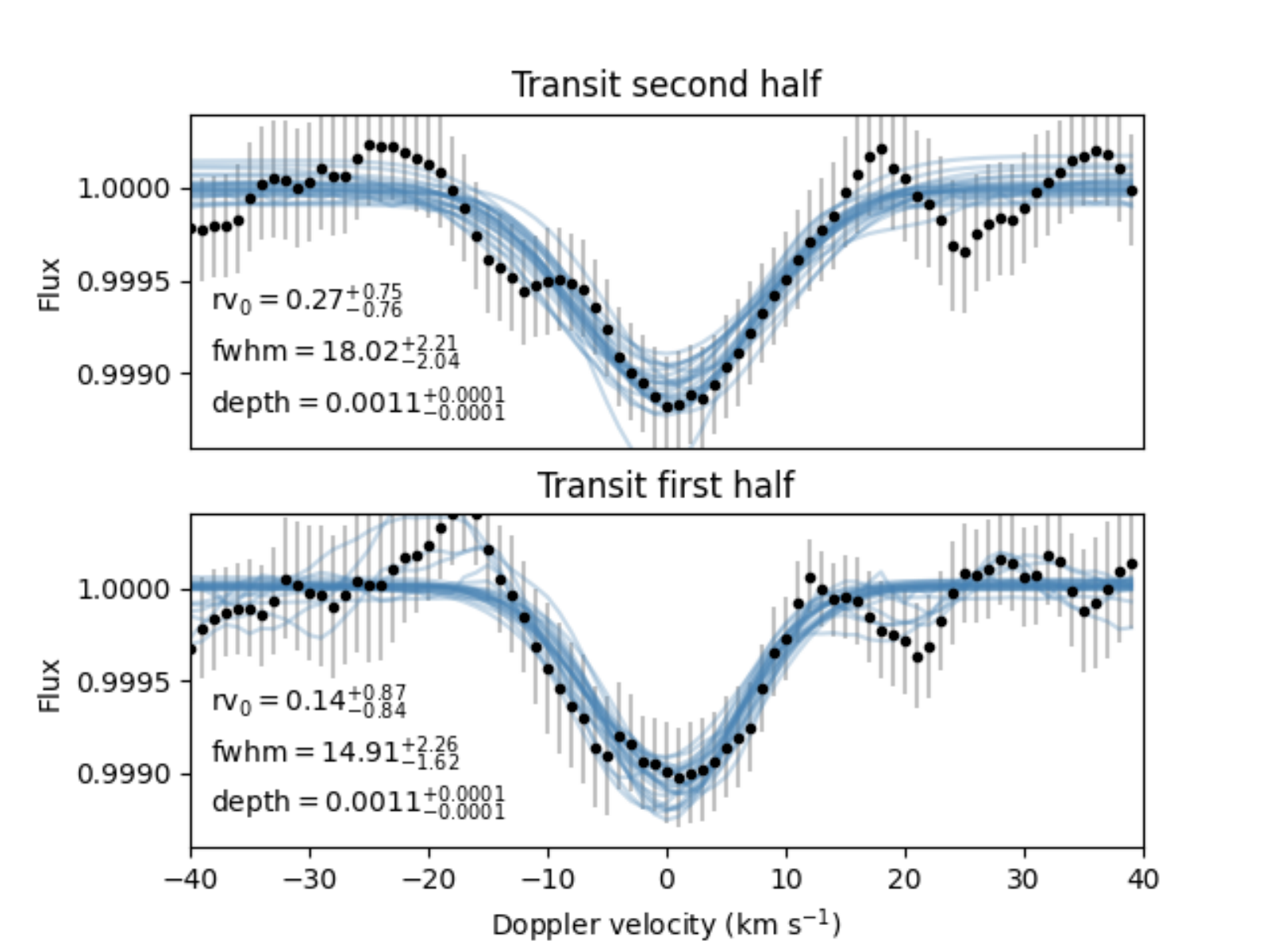}
    \includegraphics[width=0.3\textwidth]{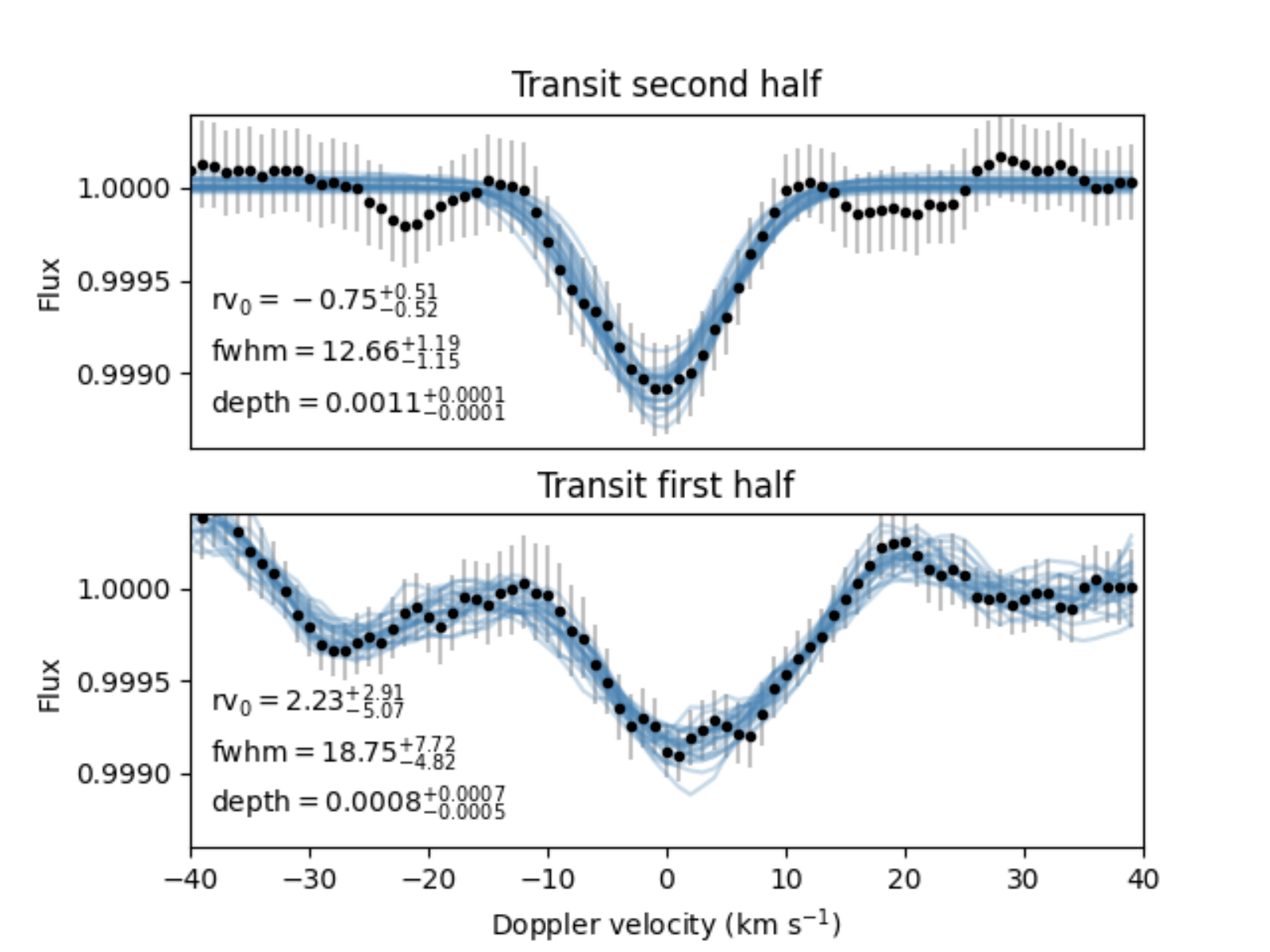}
    \includegraphics[width=0.3\textwidth]{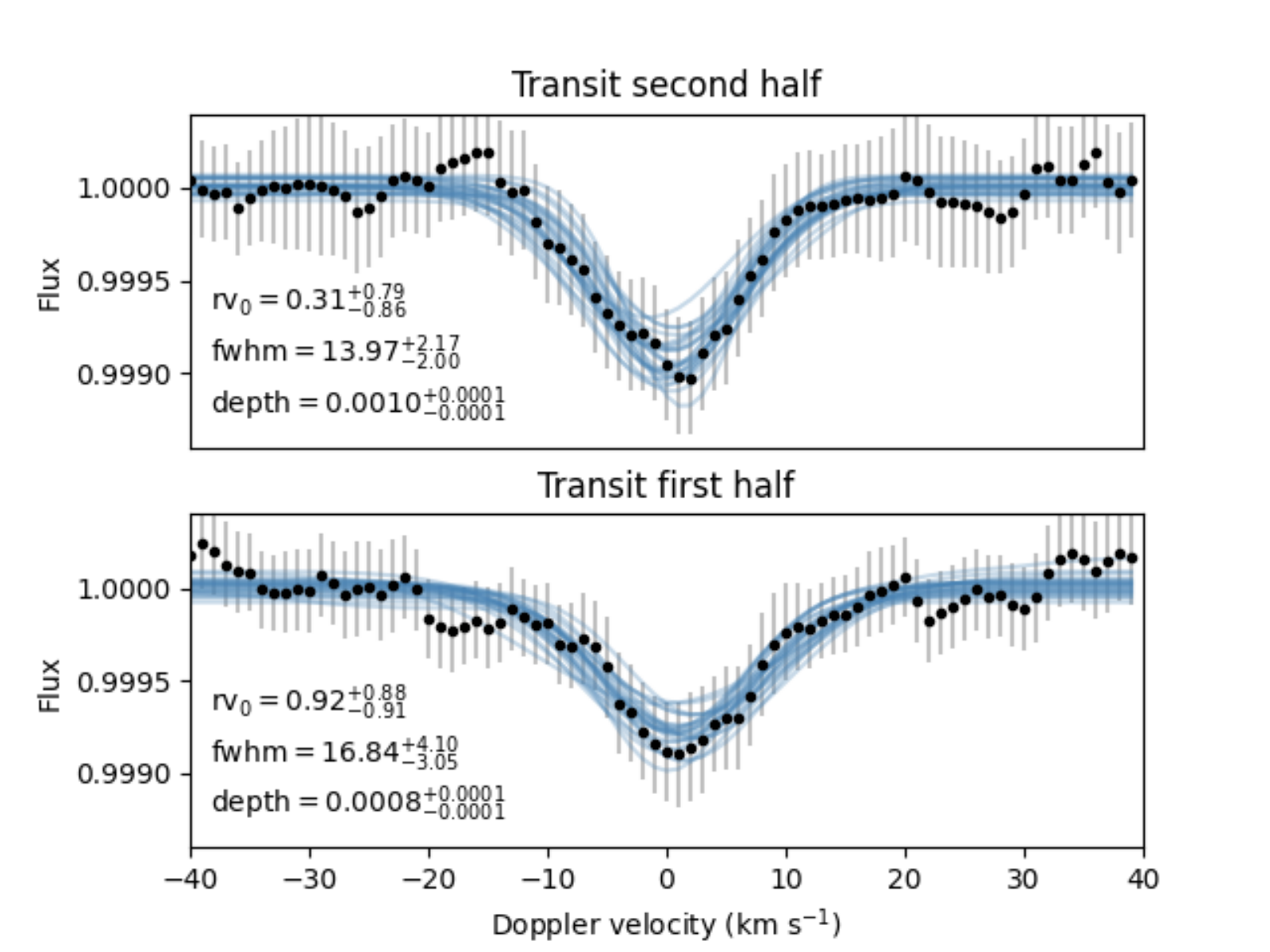}
    \includegraphics[width=0.3\textwidth]{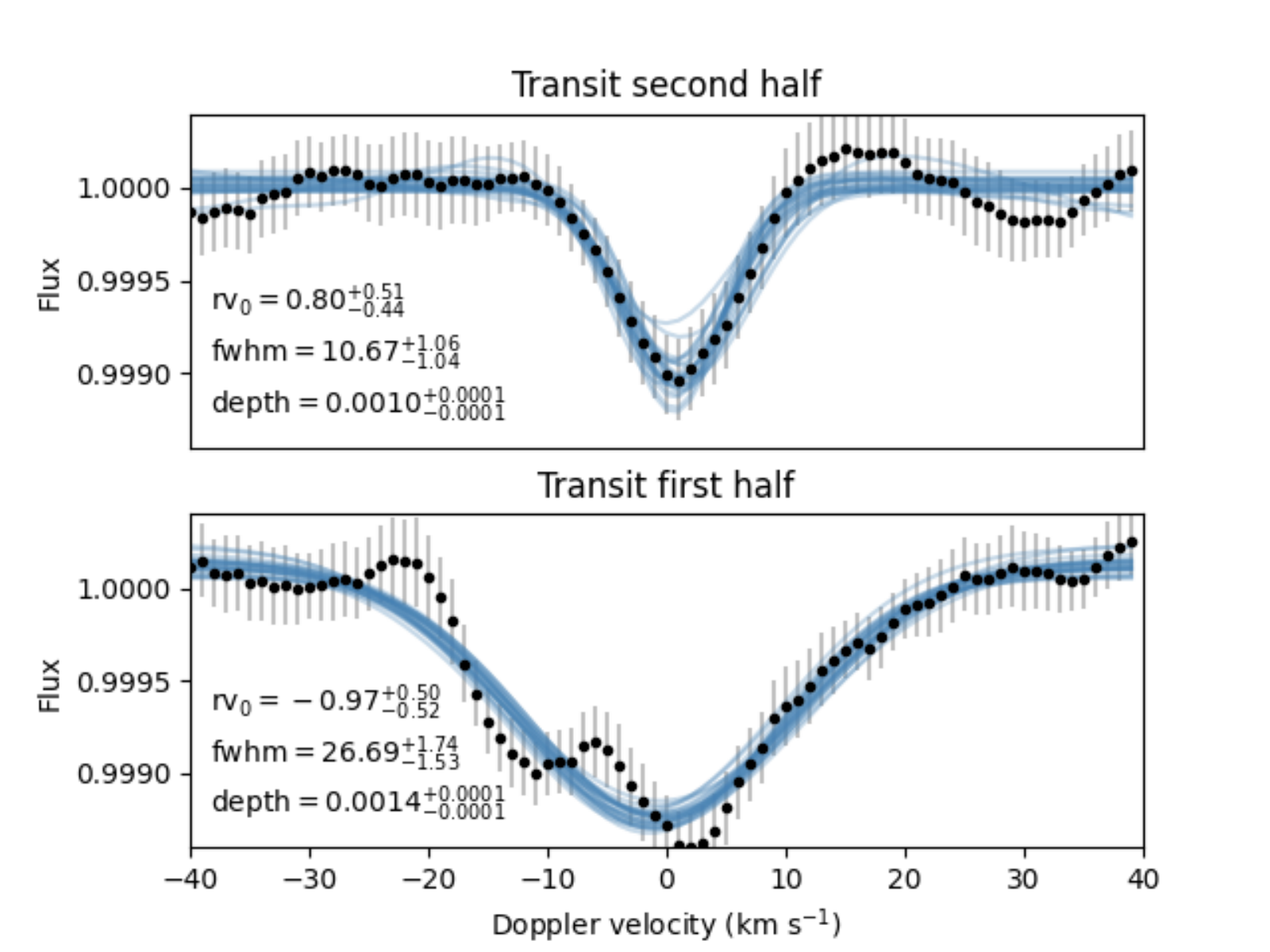}
    \includegraphics[width=0.3\textwidth]{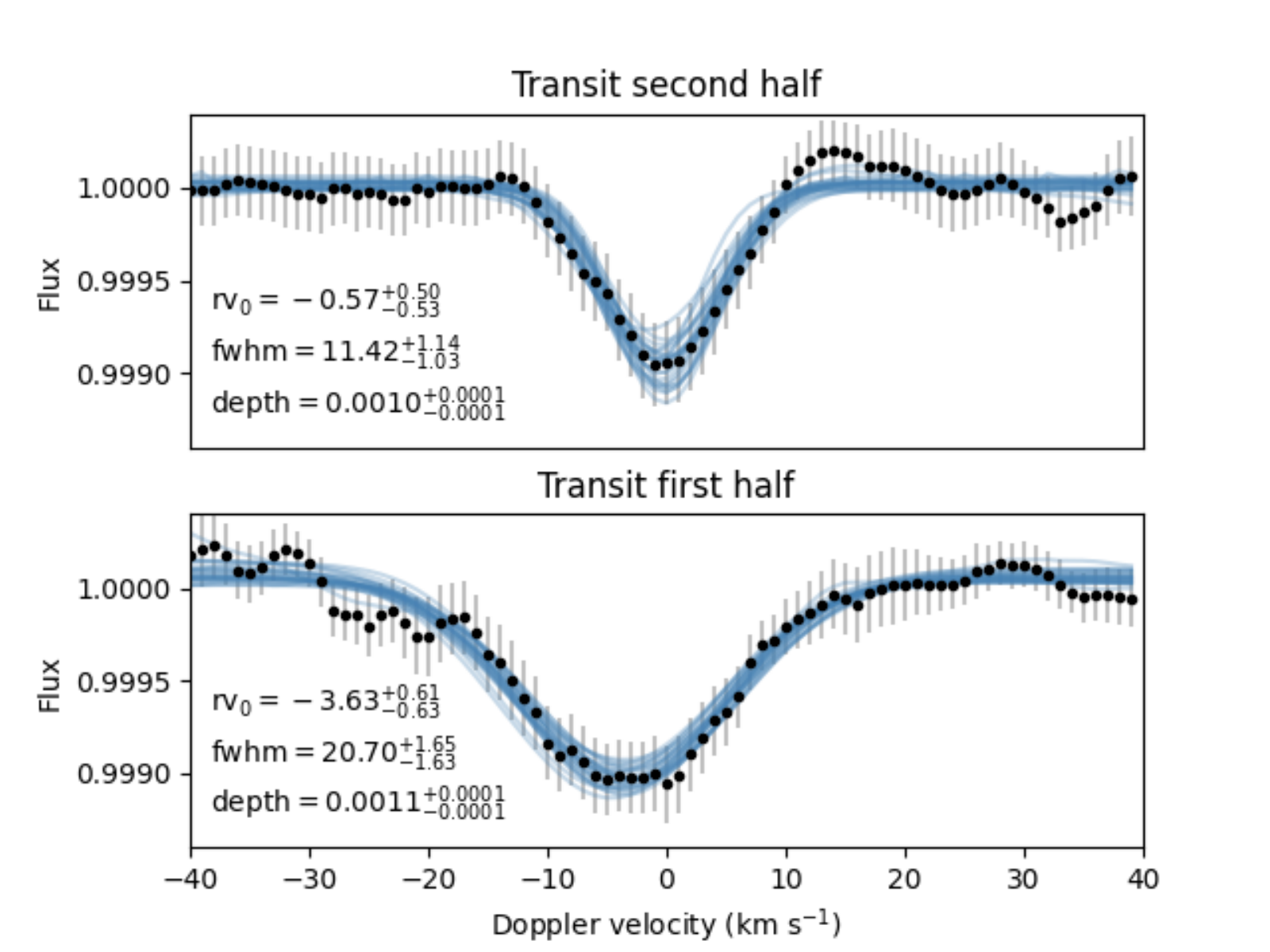}
    \caption{Averaged atmospheric signal found in the line profile residuals in the first (phase [-0.02:0.0]) and second part (phase [0.0:0.02]) of each transit. From top left to bottom right, the graphics show the results for the transits 1, 2, 3, 4, and 5.}
    \label{fig:atmo_two_halves_nights}
\end{figure*}

We tried to study the atmospheric trace behaviour in each transit, but the S/N was too low to allow us to follow the finer variations already difficult to see in Fig.~\ref{fig:atmo_results}. However, we were able to determine the overall variations between the first and second half of the transits, and we found interesting results (see Fig.~\ref{fig:atmo_two_halves_nights}).
The signal's depth is stable in all transits aside from night 4, where it decreases from the first to the second half of the transit. The FWHM show variations around 3$\sigma$ level in the last two transits (4 and 5), and just above 1$\sigma$ during transit 2: in all these three nights, the FWHM decreases from the first to the second half of the transit. It may be interesting to note that the FWHM show hints of decreasing also in transit 3, and of increasing in transit 1, but in both cases the variations are not statistically significant. Lastly, the RVs variations are visible only in the last two transits (4 and 5), where the atmospheric signal shows a small but significant redshift.

Even if the FWHM decrease found in our data does not repeat in the same way and with the same strength in every transit, it is still worthwhile to investigate some possible causes of this behaviour.
It is interesting to note that a significant variation of FWHM has been found between the elements detected in the atmosphere of KELT-20b by \cite{Hoeijmakers2020}, varying from 5.31$\pm$0.99 \kms\ for \texttt{CrII} to 33.45$\pm$3.30 \kms\ for \texttt{MgI}. Because our planetary atmospheric trace is obtained through the use of a stellar mask (where several different elements are combined), one possible interpretation for our FWHM variations could arise from a variable contribution of the chemical elements during the transit, due for example to temperature variations that may cause some of them to condense. Still, because \texttt{FeI} and \texttt{FeII} constitute more than half of the mask's lines, another possible cause may be the condensation of iron, similarly to what happens in the UHJ WASP-76b \citep{Ehrenreich2020}. 
Another interpretation could be the presence of more turbulent atmospheric conditions in the first part of the transit, due for example to a day-to-night side wind, as has been found also in both WASP-76b \citep{Ehrenreich2020} and WASP-121b \citep{Bourrier2020}.

We stress here that the atmospheric RV variations of KELT-20b are not visible when studying only the first and second half of the transit, even when combining all five transits data (see Fig.~\ref{fig:atmo_two_halves} and Fig.~\ref{fig:atmo_two_halves_nights}), but they are slightly more visible when tracing the finer atmospheric variations (see Fig.~\ref{fig:fit_atmo_trace} and Fig.~\ref{fig:atmo_results}). Due to the faintness of the signal, this study is possible only in the combined data. 


\section{Cross-correlation with FeI models}\label{sec:ccf_fe}

\cite{Nugroho2020} detected several elements in the atmosphere of KELT-20b through the CCF method, and they found a peculiar double-peak shape in the $K_p - \Delta V$ maps of \texttt{FeI}. The peaks have similar $K_p$, and they are roughly $\approx$ 10 \kms\ apart, with the secondary blueshifted peak weaker than the primary one. They reconstructed the observed structure by simulating two \texttt{FeI} signals with different amplitudes and $\Delta V$, and found the best match when masking the weaker signal (at $\Delta V$ = -10 \kms) from phase -0.01 and -0.016, simulating a delay in its appearance. This behaviour closely resembled what we observed in the planetary atmospheric RV variations, with a blueshifted signal in the final part of the transit (see Fig.~\ref{fig:atmo_results}).

Since they used the same HARPS-N observations for our first three transits, and we have two additional HARPS-N nights that were not used in their work, we decided to look for the same double-peak signature in our data. We could not use directly our LSD results, because the stellar mask contains several different elements aside from \texttt{FeI}, so we decided to apply the CCF method to obtain our own $K_p - \Delta V$ maps.

To create our model, we employed the $^\pi \eta$ line-by-line radiative transfer code \citep{Ehrenreich2006, Ehrenreich2012, Pino2018a}. This code was already used for the simultaneous interpretation of HARPS high-resolution spectroscopic observations and HST WFC3 observations \citep{Pino2018a}. For this paper, we updated the code to include:
\begin{enumerate}
\item[1)] line opacities from \texttt{FeI} and a continuum by
\texttt{H}$^-$ following \cite{Pino2020}. The \texttt{FeI} lines were taken from the VALD3 database\footnote{See \cite{Pino2020}) for a full list of references for the case of \texttt{FeI}.}, and modelled as Voigt profiles, accounting for thermal and natural broadening.\\
\item[2)] equilibrium chemistry calculations for \texttt{FeI} and \texttt{H}$^-$, to calculate their volume mixing ratios throughout the atmosphere. We employed the publicly available \texttt{FastChem} code version 2 \citep{Stock2018}.
\end{enumerate}
We employed a fixed temperature profile from \cite{Lothringer2019}, representative for a 
$T_\mathrm{eq}=2250~\mathrm{K}$ planet orbiting around an F0-type star ($T_\star = 7200~\mathrm{K}$). The other parameters employed in our model are $R_\star$,
$M_p$, $\log \mathrm{VMR_{Fe}}$, and $R_p$ at a reference pressure level of 10 bar (see Table~\ref{table:system}). \cite{Nugroho2020} demonstrated that the neutral iron lines in KELT-20b can be well represented with a hydrostatic equilibrium model, provided that the model accounts for a scale factor ($\alpha$ in their notation). Our cross-correlation scheme is not sensitive to such a scale factor, which is thus fixed to 1.

Because we are now investigating directly the planetary signal, and we are not interested anymore in the stellar contribution used to study the classical and atmospheric RML, we prepared our data in a different way than for the LSD analysis. We started again from the whole spectra, including the regions that we cut out in Sec.~\ref{sec:lsd}. We then removed the out-of-transit stellar master and telluric contamination, and we performed a cross-correlation in the stellar restframe between the data and our model on each residual spectrum \citep[as in][]{borsa2021}.
Our CCFs are defined as in Eq.~\ref{equazccf}:
\begin{equation}
CCF(v,t)= \sum\limits_{i=1}^N x_i(t,v) M_i
\label{equazccf}
\end{equation}
where $x$ are the $N$ wavelengths of the spectra taken at the time $t$ and shifted at the velocity $v$, and $M$ is the model normalised to unity. 
We impose all the model values smaller than 5\% of the maximum absorption line in the considered wavelength range to be at zero \citep[e.g.,][]{Hoeijmakers2019}.

We selected a step of 1 \kms\ and a velocity range [-200,200] \kms. The spectra are divided in segments of 200 $\AA$ \citep[e.g.,][]{Hoeijmakers2019}, then the cross-correlation is performed for each segment. 
We masked the wavelength range 5240-5280 \AA, which is heavily affected by telluric contamination. Then for each exposure we applied a weighted average between the CCFs of the single segments, 
where the weights applied to each segment are the sum of the depths of the lines in the model and the inverse of the standard deviation of the segment (i.e., the higher the S/N, the larger the weight).
We then averaged all the in-transit CCFs after shifting them in the planetary restframe, for a range of $K_p$ values from 0 to 300 \kms, in steps of 1 \kms.  The shift is performed by subtracting the planetary RV calculated for each spectrum as $v_{p}=K_p \times \sin{2\pi \phi}$, where $\phi$ is the orbital phase. 
As a last step, we subtracted the \vsys\ from all the averaged CCFs to obtain the $K_p - \Delta V$ maps.
We then created S/N maps by computing the standard deviation of each $K_p - \Delta V$ map far from the planetary signal (i.e. excluding the region from \vsys\ = -40 \kms\ to \vsys\ = 40 \kms), and then dividing the $K_p - \Delta V$ maps by these values.

\begin{figure}
    \centering
    \includegraphics[trim=0 0 0 38, clip, width=\columnwidth]{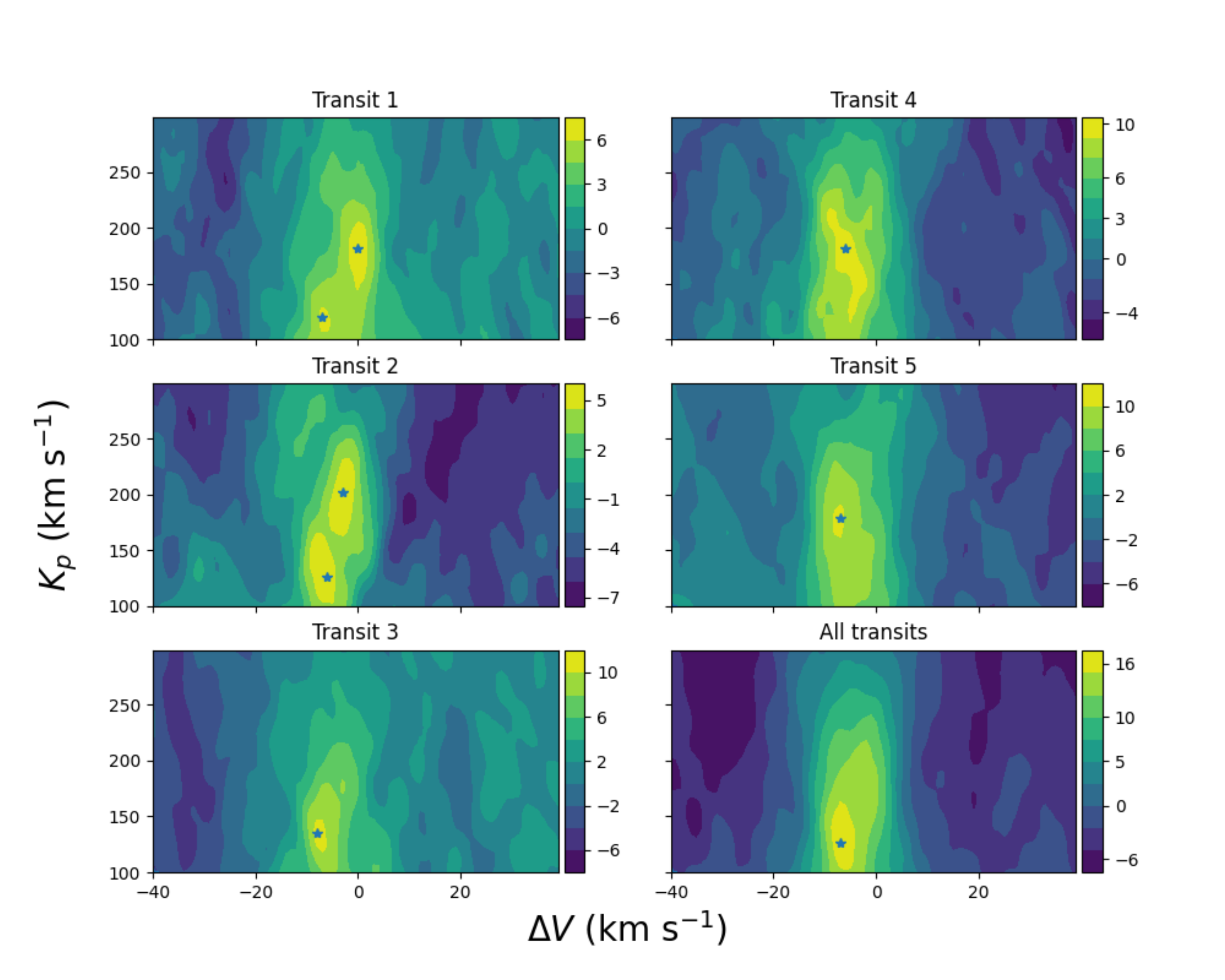}
    \caption{Contour plots of the $K_p - \Delta V$ maps obtained from the cross-correlation of the residual spectra with the planetary \texttt{FeI} model. All the five transit nights and the combined data (bottom right panel) are shown. The prominent peaks are indicated with star symbols.}
    \label{fig:kp}
\end{figure}

Our results are shown in Fig.~\ref{fig:kp}: we identified the strongest peak or peaks in each map with a local maxima algorithm. We found the double-peak feature quite clearly in the first two transits, while only the weaker blueshifted signal is present in transit 3. We remind that those are the same data analysed by \cite{Nugroho2020}. In transits 4 and 5 only the stronger signal is visible.

To further investigate the variability of the \texttt{FeI} signal as a function of transit, we ran MCMC simulations via the Python \texttt{emcee} package and drove their evolution via the likelihood scheme of \citet{Brogi2019}.
The CCF defined in Eq.~\ref{equazccf}, and commonly used in stellar radial velocities, is actually called cross-covariance in statistics. Compared to the statistical cross-correlation, it misses a normalisation factor that would force the CCF between -1 (perfect anti-correlation) and +1 (perfect correlation). The quantity in Eq.~\ref{equazccf} is thus exactly the same as the cross-covariance $R$ defined in Eq.~9 of \citet{Brogi2019} and used for the MCMC simulations. However, the likelihood function of \citet{Brogi2019} contains additional terms, namely the data and model variances, which give information about the shape and amplitude of spectral lines.
We also note that that the cross-correlation maps in Fig.~\ref{fig:kp} are converted in S/N by dividing through the standard deviation far from the peak: using the full statistical formula for cross-correlation we would get the same S/N, except for second-order variations of the model variance as a function of \vsys\ and $K_p$.

We chose to present here both analyses because the CCF approach is less sensitive to the modelling and thus more appropriate to capture the initial detection of \texttt{FeI} even with an approximate template or a binary mask. The CCFs can also be compared with the existing literature. However, the likelihood approach allows us to better constrain the parameter space and explore the statistical evidence for night-to-night variability, and thus we opted for presenting both analyses.

\begin{figure}
    \centering
    \includegraphics[width=\columnwidth]{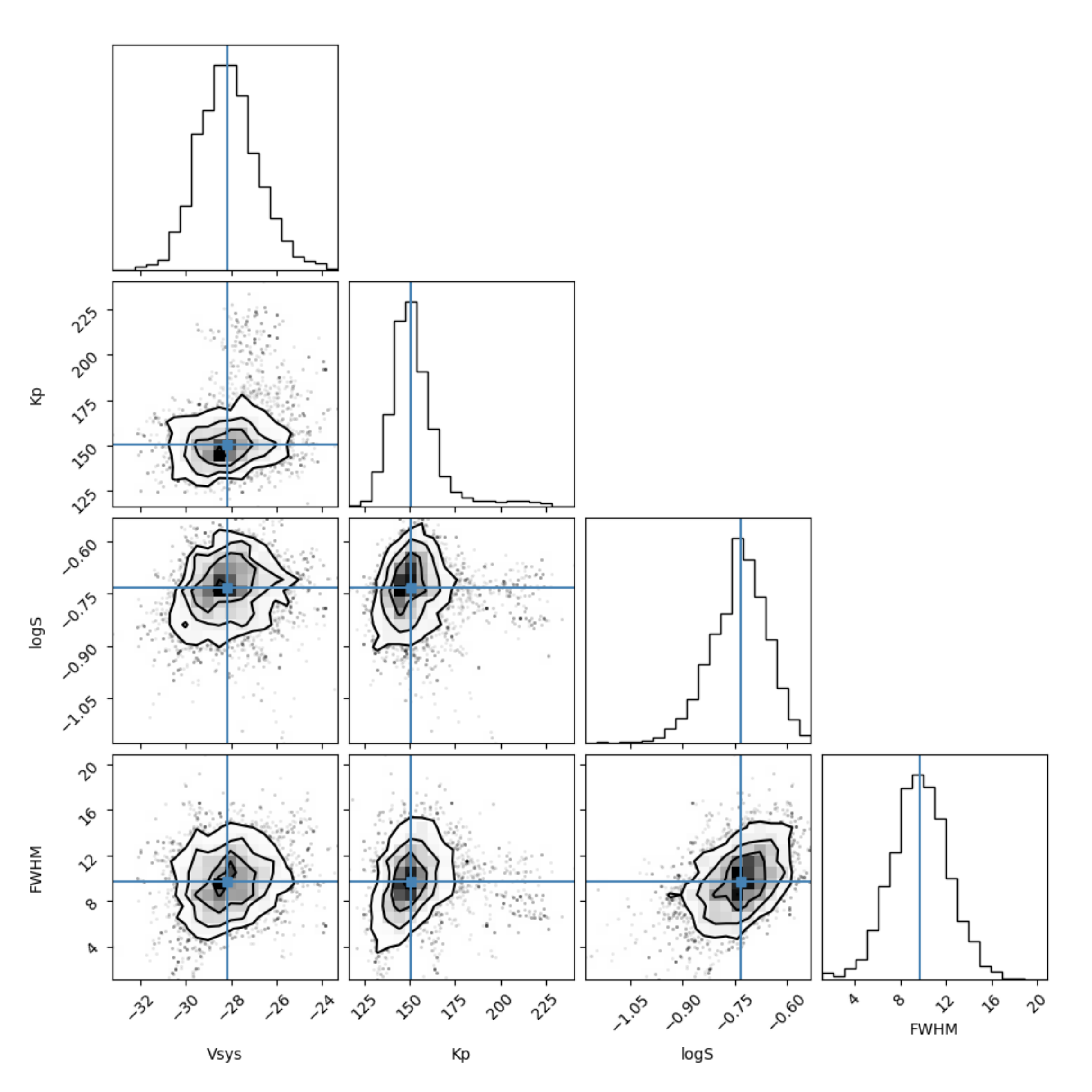}
    \caption{Corner plot of the posterior distribution found with MCMC simulations and \citet{Brogi2019} likelihood for the transit 2, where the double-peak structure is clearer in the $K_p - \Delta V$ contour maps. No evidence of such structure is found here.}
    \label{fig:corner}
\end{figure}

In our simulations, the likelihood was maximised as a function of four parameters: the two velocities (orbital and systemic), the FWHM of the line profile, and the logarithm of a scaling factor, $\log S$. While the measured systemic velocity is consistent within 1$\sigma$ between the nights (aside from the first night), the other three parameters show a clear variability, as reported in Table~\ref{table:kp_vsys}. Here we redefine $K_p$ as $K_\mathrm{p+atmo}$, because it combines both the projected orbital velocity of the planet and a contribution from the atmosphere's physics and dynamics.
With this more refined method we found no evidence of the double-peak structure: in Fig.~\ref{fig:corner} we show the corner plot of the posterior distribution found for the transit night 2, where the double-peak structure is instead clearly visible in Fig.~\ref{fig:kp}.
We also investigated the combined data by fixing the FWHM and $\log S$ to their best-fitting values for each night, and then running an additional MCMC with the five nights combined. Also this further test did not confirm a possible double solution. The posterior in $K_\mathrm{p+atmo}$ appears instead single-peaked, and centered at an intermediate value of 142 \kms.
The new results show instead that the values of $K_\mathrm{p+atmo}$ and $\log S$ are variable over some of the transits. Only the last two transits (4 and 5) yield a $K_\mathrm{p+atmo}$ value fully compatible with the theoretical $K_p$ estimated in Sec.~\ref{sec:atmo_trace}, while in the other transits and the combined data the $K_\mathrm{p+atmo}$ is systematically lower than the theoretical $169\pm6$~\kms\ value.

These results hint to the fact that the dynamics probed by the \texttt{FeI} signal, as well as the overall strength of the iron lines, may both change from transit to transit. Furthermore, there is some evidence that the broadening of the line profile also varies, as shown by the retrieved values of FWHM, even if these variations are not strongly statistically significant. Still we note here that, comparing these results with those of Sec.~\ref{sec:atmo_trace}, the nights with larger $K_\mathrm{p+atmo}$ (4 and 5) are those with the larger FWHM variations of the atmospheric trace between the two half of the transit, while the night 1 is the one with the more deviant values, and also the only one where the atmospheric trace shows a possible increase of the FWHM from the first to the second half of the transit (see Fig.~\ref{fig:atmo_two_halves_nights}). The difference between the values of FWHM found here and those of the atmospheric trace may arise from the fact that the atmospheric trace described in Sec.~\ref{sec:atmo_trace} is caused by a combination of the different elements found in the stellar mask. Additionally, we confirm the blueshift of the \texttt{FeI} signal found also in literature \citep{Stangret2020,Hoeijmakers2020,Nugroho2020},

\begin{table}
\caption{Signal position, and width and model's scale factor found with the MCMC simulations and the likelihood from \citet{Brogi2019}}
\label{table:kp_vsys} 
\centering 
\begin{tabular}{c c c c c}  
\hline\hline  
Night & $K_p$ ($K_\mathrm{p+atmo}$) & $\Delta V$ & FWHM & $\log S$ \\  
 & (\kms) & (\kms) & (\kms) & \\  

\hline  
   1 & 99.5$^{+13.6}_{-20.2}$ & -1.0$^{+1.7}_{-1.4}$ & 16.7$^{+4.2}_{-2.8}$ & -0.20$^{+0.05}_{-0.05}$ \\ 
   2 & 150$^{+13.0}_{-9.1}$ & -3.7$^{+1.4}_{-1.3}$ & -9.7$^{+2.5}_{-2.6}$ & -0.73$^{+0.08}_{-0.09}$ \\
   3 & 129.7$^{+4.4}_{-7.2}$ & -6.0$^{+0.7}_{-0.6}$ & -6.4$^{+3.8}_{-1.9}$ & -0.55$^{+0.07}_{-0.06}$ \\
   4 & 163.1$^{+21.3}_{-15.2}$ & -3.3$^{+0.8}_{-0.9}$ & -12.7$^{+1.8}_{-1.8}$ & -0.40$^{+0.04}_{-0.05}$ \\
   5 & 161.1$^{+4.9}_{-4.9}$ & -4.9$^{+0.5}_{-0.5}$ & -9.7$^{+1.2}_{-1.2}$ & -0.33$^{+0.03}_{-0.03}$ \\ 
   All data & 142$^{+7}_{-6}$ & -4.7$^{+0.3}_{-0.3}$ &  & \\
\hline       
\end{tabular}
\end{table}


\section{Conclusions} \label{sec:conclusion}
Because of its high $T_\mathrm{eq}$ (2260 $\pm$ 50 K), the atmospheric spectrum of the ultra-hot Jupiter KELT-20b correlates with the stellar mask used to compute the host star mean line profiles. We were thus able to detect and characterise the atmospheric RML effect present in the stellar RV time-series, which resulted in an estimation of the size of the planetary atmosphere that correlates with the mask ($R_{p+atmo}/R_p = 1.13 \pm 0.02$). This is in agreement with literature values from metal line-depths, confirming the reliability of the atmospheric RML method.
In addition to that, we could isolate the atmospheric trace in the mean line profile tomography: the high-resolution, high S/N of our data allowed us to fit the atmospheric signal and follow its variations during the transit.

We found possible variations of RV, FWHM and depth of the atmospheric signal during the combined transit data and a different behaviour during different transits.
There is a decrease of the FWHM during two transits, and just above 1$\sigma$ level during a third one, while in another transit and in the combined data the measured decrease may be present, but it is not statistically significant. This possible greater FWHM spread during the first part of the transit may hint at turbulent conditions, that become more stable in the second part of the transit. This behaviour resembles that found by \cite{Ehrenreich2020} in WASP-76b, and \cite{Bourrier2020} in WASP-121b, and confirms the existence of different structures between morning and evening terminators, as suggested by \cite{Hoeijmakers2020}: in their work, they analysed only one transit and so they could not exclude a spurious nature for the RV variability.
Another possible interpretation for the FWHM variations may be the variable contribution of elements to the overall atmospheric signal during the transit: because the stellar mask contains different elements, all of them contribute to the resulting atmospheric trace, but their relative abundances may change during the transit due to, for example, temperature variations that may cause some of them to condense. This may result in a FWHM variation due to the significant FWHM differences between the elements detected in KELT-20b atmosphere \citep{Hoeijmakers2020}. Seeing as more than half of the stellar mask lines are \texttt{FeI} and \texttt{FeII} line, the condensation of iron \citep{Ehrenreich2020} may play a role in this situation.

The RV variations are more visible in the combined data of the atmospheric trace, that confirms the presence of atmospheric dynamics with a blueshift of the signal during egress.
These RV variations led us to explore the results from \cite{Nugroho2020}, who found a double-peak feature in their \texttt{FeI} $K_p - \Delta V$ maps. This feature consisted of a primary peak at $\Delta V$ = 0 \kms\ and a weaker secondary peak at $\Delta V$ = -10 \kms. Their best match with simulated signals indicated a delayed appearance of the weaker blueshifted signal, which agrees well with the blueshift we found in the second part of the transit.
The atmospheric trace that we studied in Sec.\ref{sec:atmo_trace} was found in the mean line profiles obtained using the LSD software with a stellar mask suited for the host star KELT-20, and as such we could not directly compare our results with those of \citet{Nugroho2020}. We decided then to create our own $K_p - \Delta V$ maps using the standard cross-correlation method with a planetary atmospheric \texttt{FeI} model. Initially we found the same double-peak structure in transits 1 and 2 with a simple contour analysis, but a more refined study using MCMC simulations driven via the likelihood scheme of \citet{Brogi2019} showed only a single significant peak per night. Nevertheless, we did find some indication of the variability of the signal from one transit to another, that aligns well with the tentative results from the line profile tomography.

To conclude, we used different methods (line profile tomography and \texttt{FeI} CCFs) to search independently for variability in the atmospheric signal of KELT-20b, and we found indication of FWHM and RV variability during some of the transits. We also confirm the blueshift of the \texttt{FeI} signal and the reliability of the atmospheric RML method to estimate the atmospheric extension.

\begin{acknowledgements}
This work has made use of the VALD3 database, operated at Uppsala University, the Institute of Astronomy RAS in Moscow, and the University of Vienna. MB acknowledges support from the UK Science and Technology Facilities Council (STFC) research grant ST/S000631/1. GSc acknowledges the funding support from Italian Space Agency (ASI) regulated by “Accordo ASI-INAF n. 2013-016-R.0 del 9 luglio 2013 e integrazione del 9 luglio 2015”. FB acknowledges support from PLATO ASI-INAF agreement n. 2015-019-R.1-2018.
The research leading to these results has received funding from the European Research Council (ERC) under the European Unions Horizon 2020 research and innovation programme (grant agreement no. 679633, Exo-Atmos). 
\end{acknowledgements}

\bibliographystyle{aa} 
\bibliography{kelt20.bib} 

\begin{appendix}
\section{Atmospheric trace analysis on dataset B}\label{app:cabot}
In order to ensure that the removal of the Doppler shadow did not unduly affect the study of the atmospheric trace, we performed the removal with two methods, which generated two datasets. While the results from dataset A are shown in the paper, we show here the same analysis performed on dataset B.

\begin{figure}
    \centering
    \includegraphics[trim=0 0 0 38, clip, width=\columnwidth]{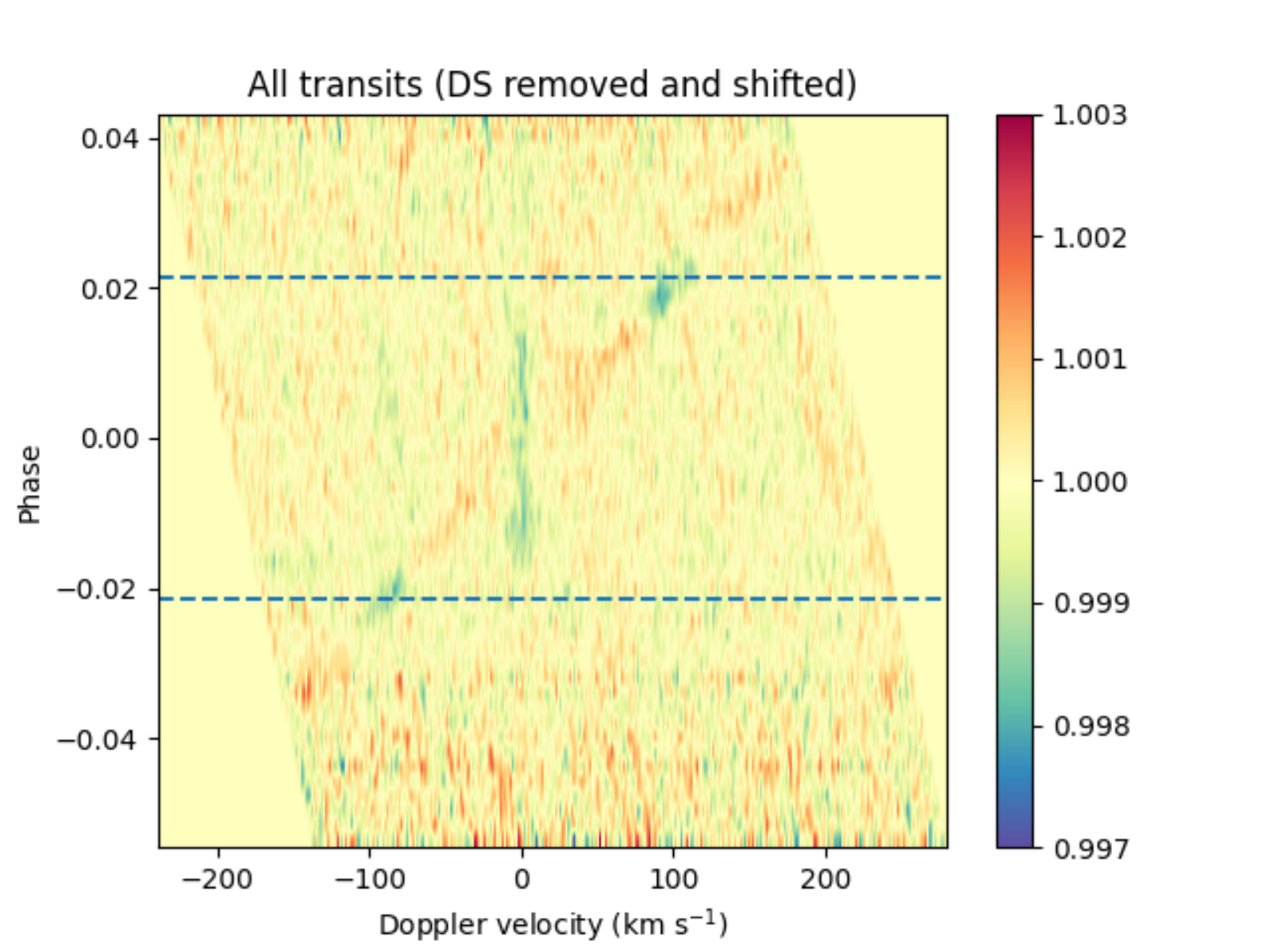}
    \caption{Dataset B: line profile residuals after the Doppler shadow's removal, and after being shifted in the planet reference frame. The atmospheric trace is clearly visible and centered around 0~\kms. There is still a visible Doppler shadow residual.}
    \label{fig:atmo_trace_cabot5}
\end{figure}

The line profile residuals after the Doppler shadow's removal are shifted in the planetary reference frame and shown in Fig.~\ref{fig:atmo_trace_cabot5}).
It is clearly evident that the Doppler shadow's removal was less efficient in this case than in dataset A (see Fig.~\ref{fig:atmo_trace}), as evidenced by the large residuals left in the tomography.
We then smoothed each in transit residual by applying a $3^{rd}$ degree Savitzky-Golay filter with a 15 pixels window, and we fitted the atmospheric signal using MCMC with a correlated noise model.

\begin{figure}
    \centering
    \includegraphics[trim=0 0 0 38, clip, width=\columnwidth]{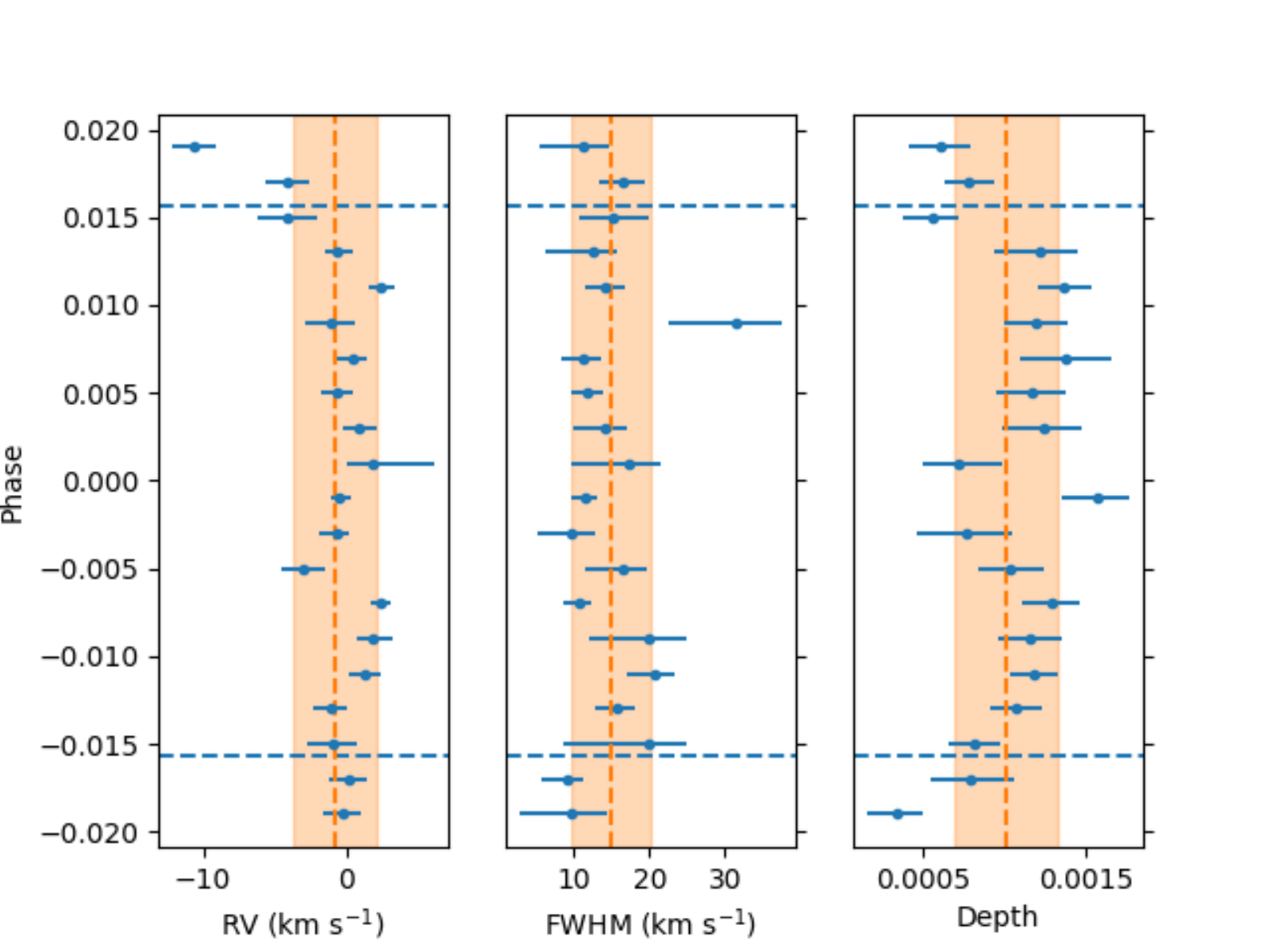}
    \caption{RVs, FWHM and depth of the atmospheric signal during transit using dataset B: the vertical dashed orange lines and shaded areas show the mean values and relative standard deviation for each of the three quantities. All the data are comprised between $t_0$ (start of ingress) and $t_2$ (end of egress), while the horizontal dashed blue lines indicate $t_2$ (end of ingress) and $t_3$ (start of egress)}
    \label{fig:atmo_results_cabot5}
\end{figure}

The resulting RVs, FWHMs and depths are shown in Fig.~\ref{fig:atmo_results_cabot5}.
As for dataset A, we found a blueshift during egress, while the FWHM is more stable (aside from the same outlier found in dataset A) and the depth shows a symmetric increase and subsequent decrease during transit.

\begin{figure}
    \centering
    \includegraphics[width=\columnwidth]{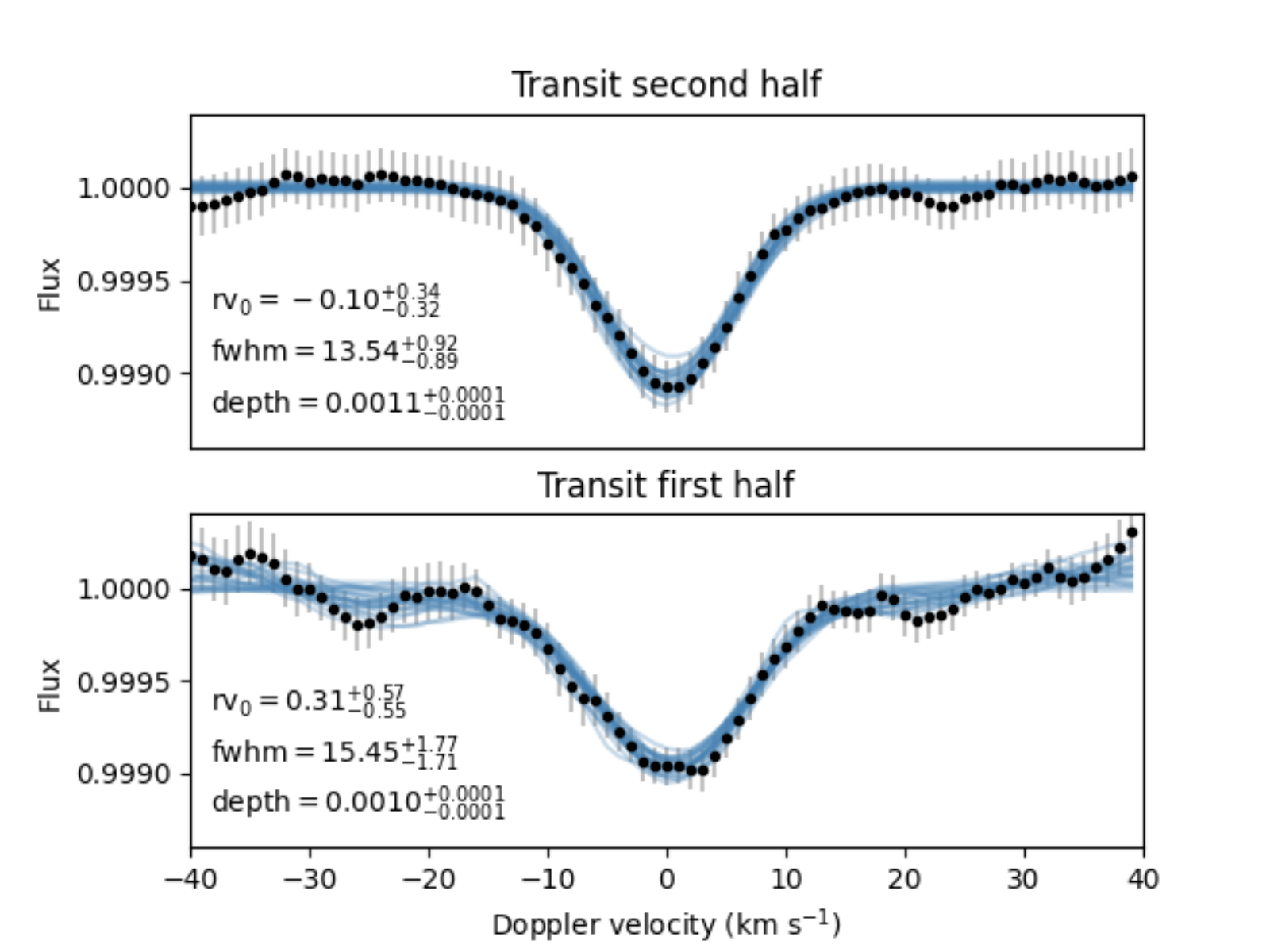}
    \caption{Averaged atmospheric signal in the first (phase [-0.02:0.0]) and second part (phase [0.0:0.02]) of the transit with relative fit using dataset B.}
    \label{fig:atmo_two_halves_cabot5}
\end{figure}

\begin{figure*}
    \centering
    \includegraphics[width=0.3\textwidth]{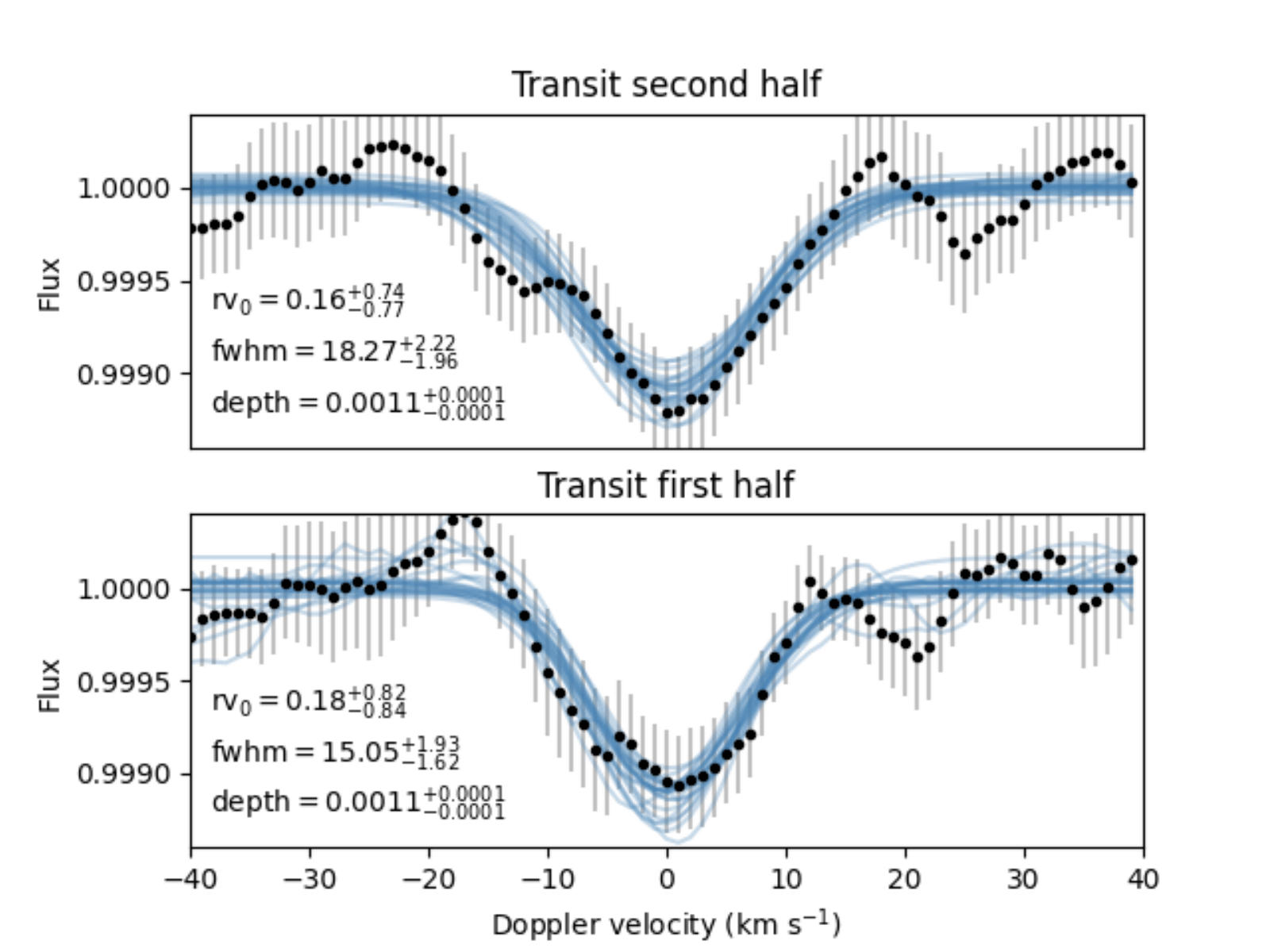}
    \includegraphics[width=0.3\textwidth]{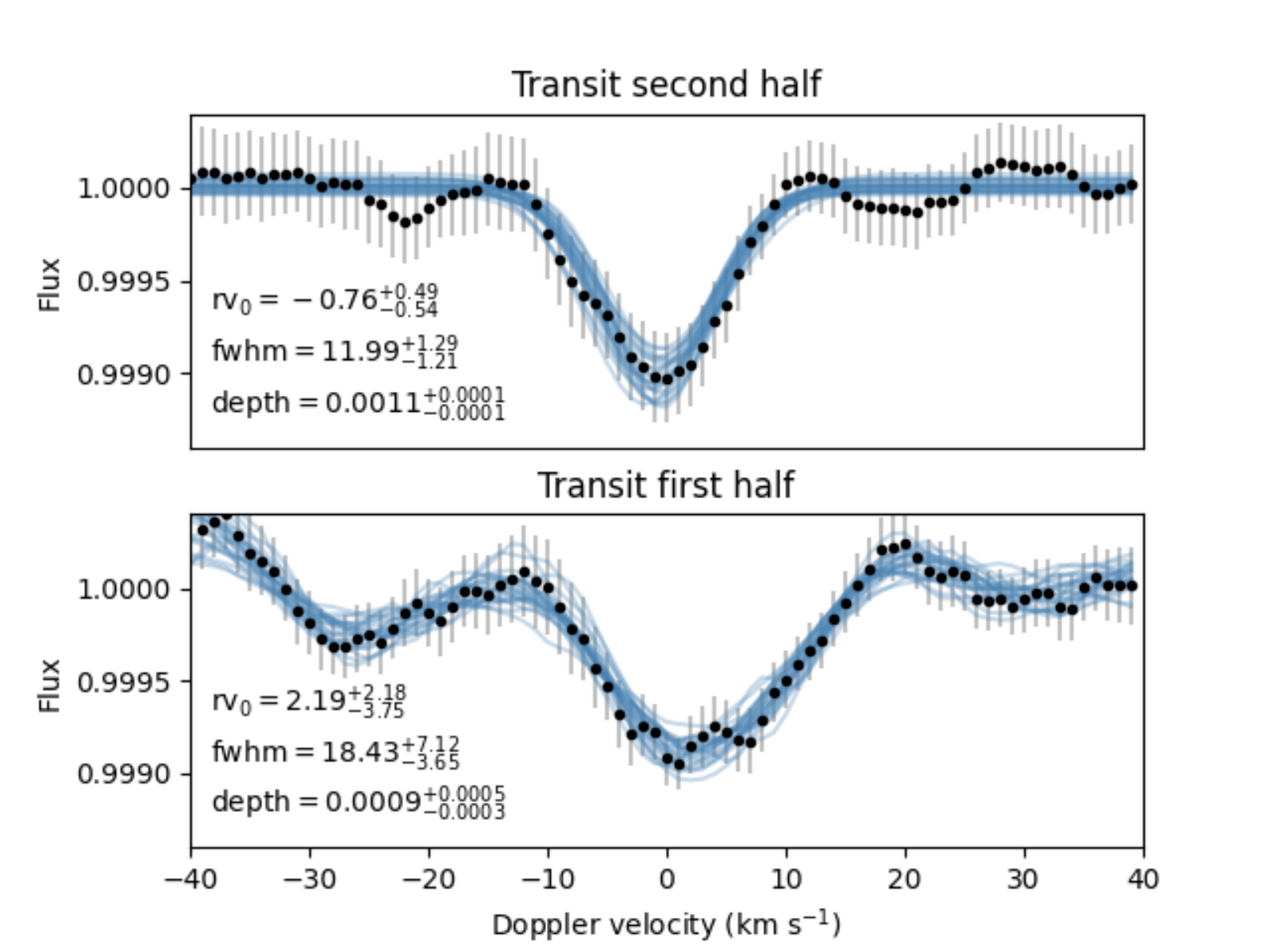}
    \includegraphics[width=0.3\textwidth]{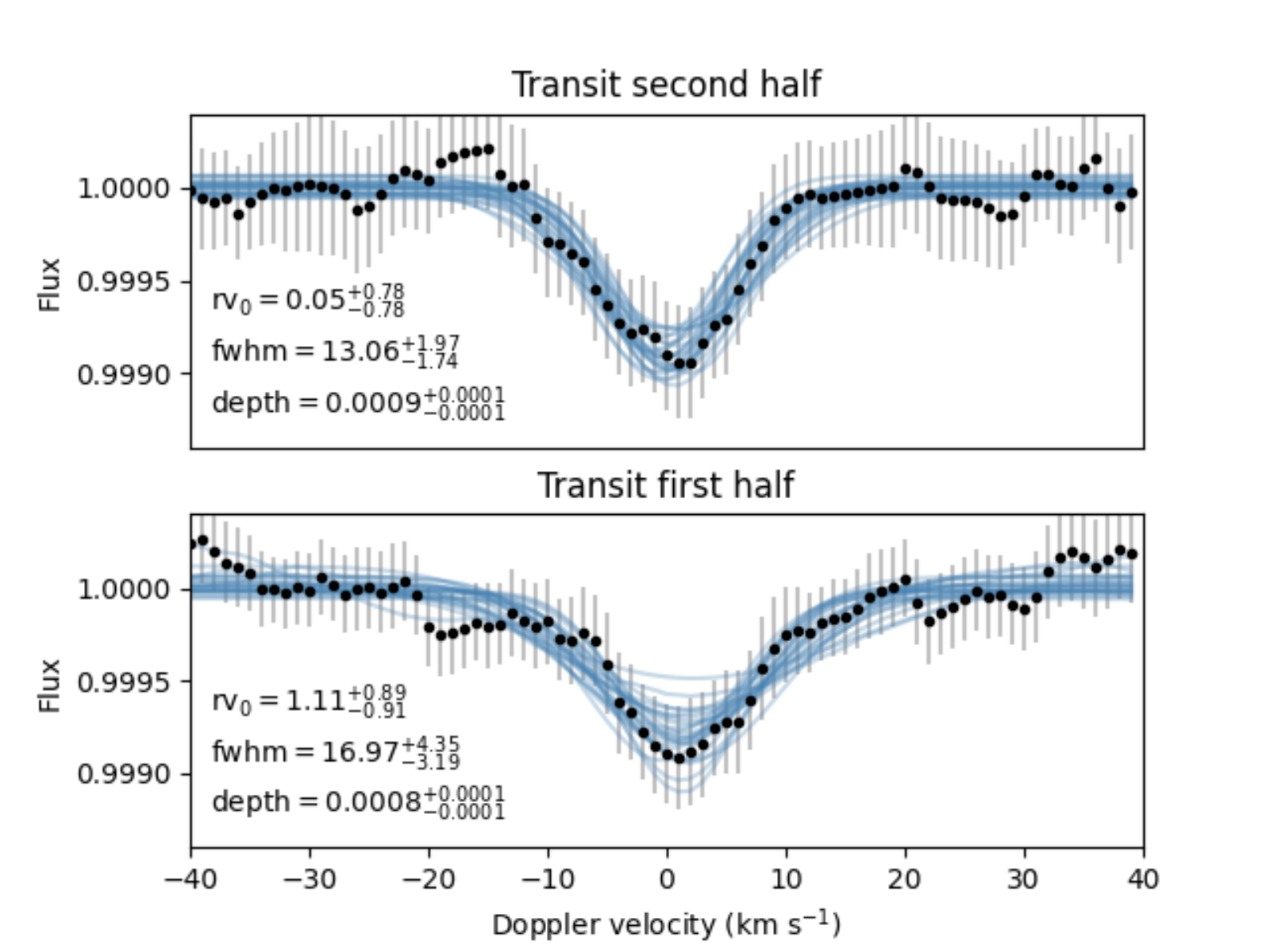}
    \includegraphics[width=0.3\textwidth]{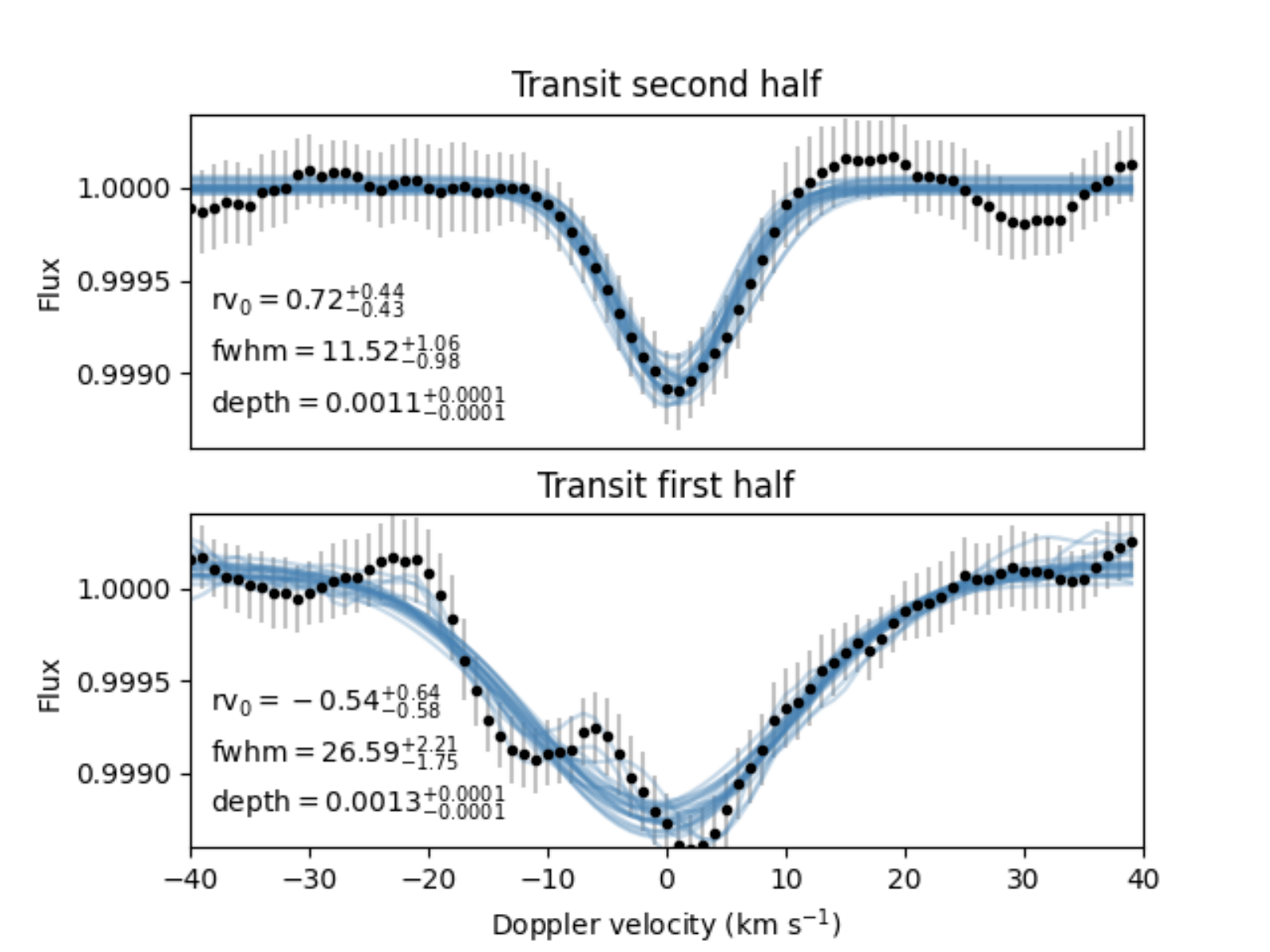}
    \includegraphics[width=0.3\textwidth]{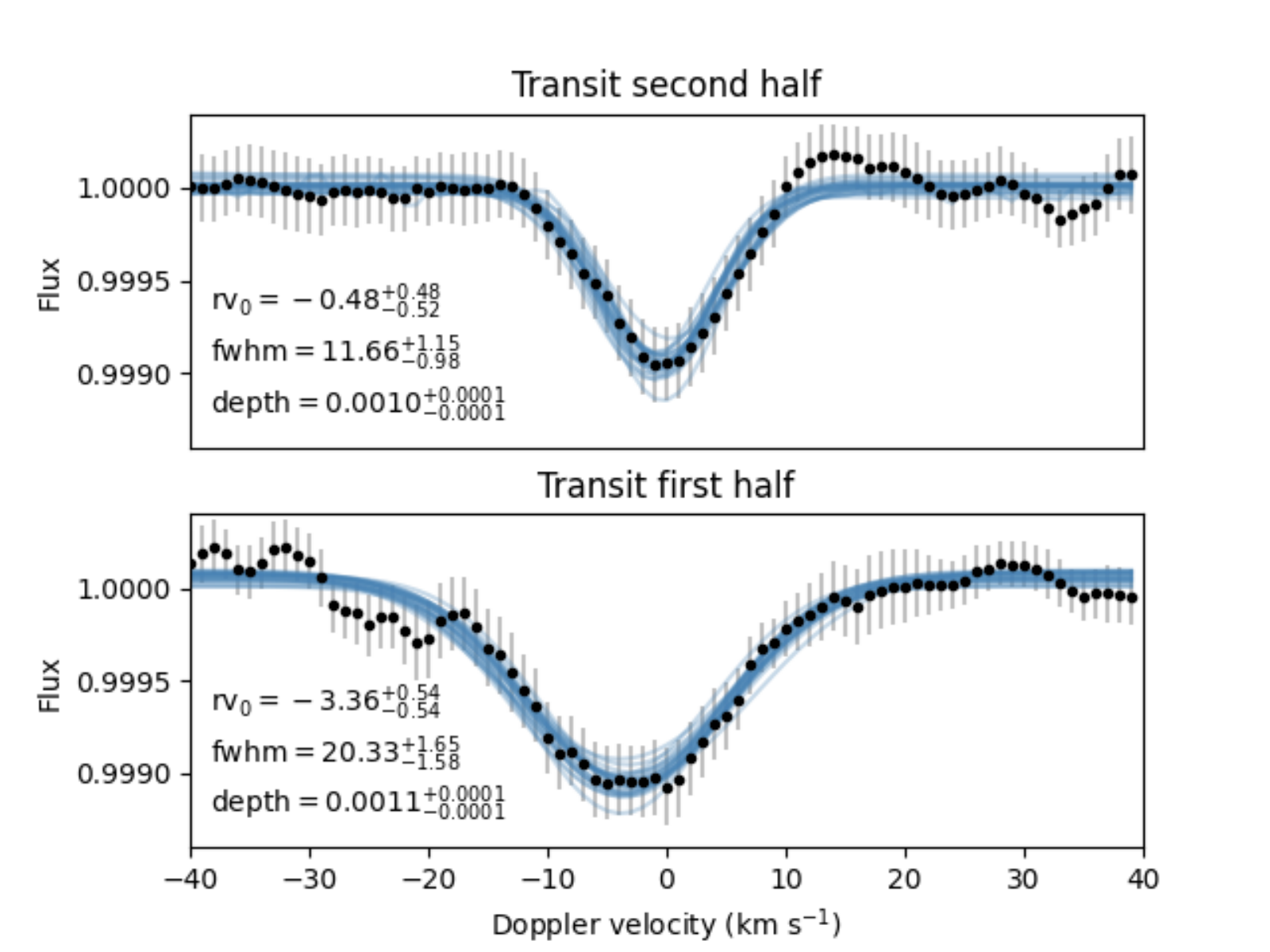}
    \caption{Averaged atmospheric signal in the first (phase [-0.02:0.0]) and second part (phase [0.0:0.02]) of each transit using dataset B. From top left to bottom right, the graphics show the results for the nights 1, 2, 3, 4, and 5.}
    \label{fig:atmo_two_halves_nights_cabot5}
\end{figure*}

We studied the overall variations during transit by averaging all residuals in the first and second half of the transit (phases [-0.02:0.0] and [0.0:0.02]), see Fig.~\ref{fig:atmo_two_halves_cabot5}.
We performed the same study also on each individual night (see Fig.~\ref{fig:atmo_two_halves_nights_cabot5}).
We found mostly the same results of dataset A: the atmospheric depth is stable in all transit nights (also during night 4 the variation falls exactly at 1$\sigma$ level), the RV variations are  statistically significant only in night 4, where a small redshift is visible, and the FWHM decreases during the transit in nights 2 (just above 1$\sigma$ level), 4, and 5 (around the 3$\sigma$ level).
The results are qualitatively identical to those obtained with dataset A, showing that the choice of the Doppler shadow removal method does not influence our work.

\section{Atmospheric trace analysis on unfiltered dataset A}\label{app:savgol}

\begin{figure}
    \centering
    \includegraphics[trim=0 0 0 38, clip, width=\columnwidth]{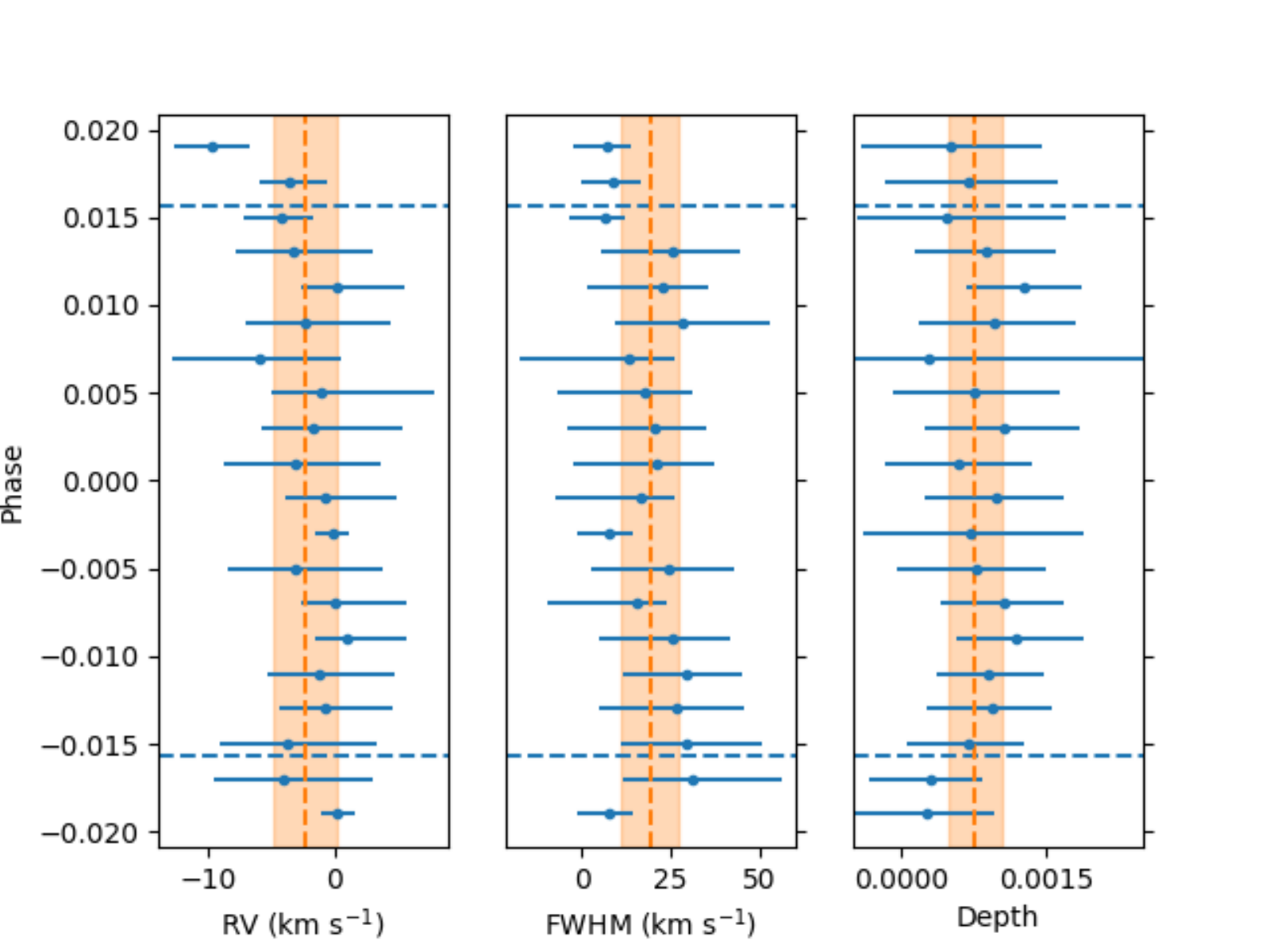}
    \caption{RVs, FWHM and depth of the atmospheric signal during transit using the unfiltered data: the vertical dashed orange lines and shaded areas show the mean values and relative standard deviation for each of the three quantities. All the data are comprised between $t_0$ (start of ingress) and $t_2$ (end of egress), while the horizontal dashed blue lines indicate $t_2$ (end of ingress) and $t_3$ (start of egress)}
    \label{fig:atmo_results_noSG}
\end{figure}

\begin{figure}
    \centering
    \includegraphics[width=\columnwidth]{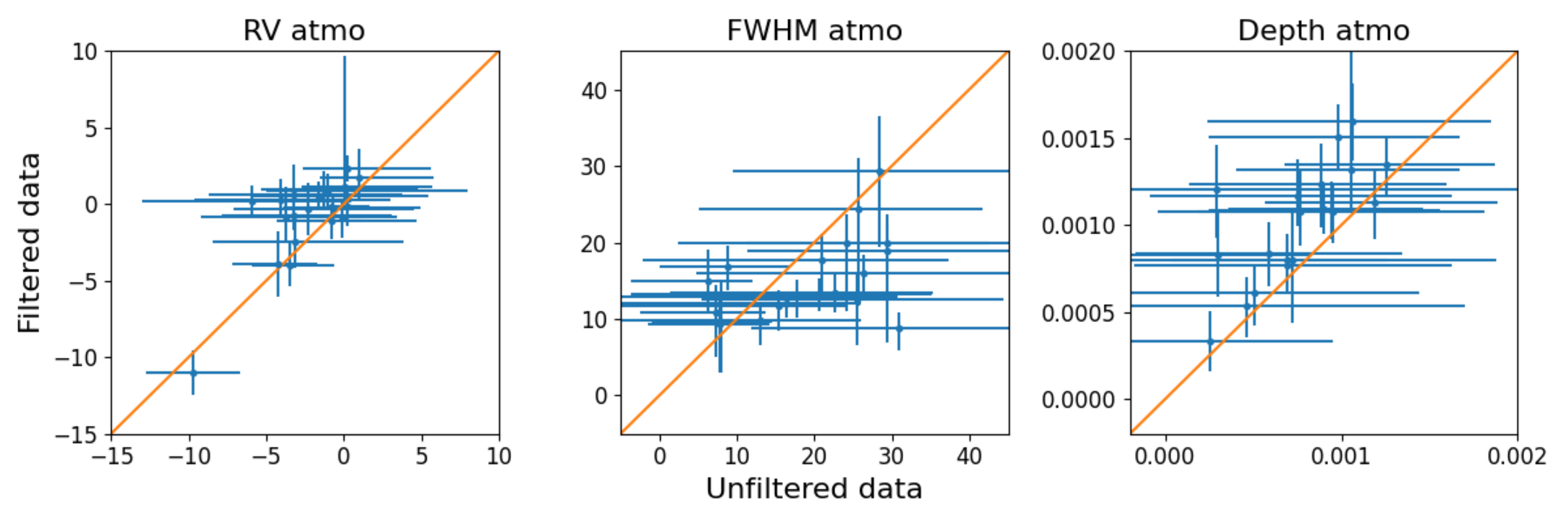}
    \caption{Comparison between the atmospheric trace values of RV, FWHM and depth found in dataset A using the unfiltered data (x-axis) and filtered data (y-axis). The orange lines show the one-to-one correlation.}
    \label{fig:atmo_correlation}
\end{figure}

We performed the same analysis shown in Sec.~\ref{sec:atmo_trace} on the unfiltered dataset A, combining all five transits together. While the atmospheric trace signal is much noisier, the overall behaviour of its FWHM, RV and depth, shown in Fig.~\ref{fig:atmo_results_noSG}, closely resembles that found in the filtered data (Fig.~\ref{fig:atmo_results}), as can be seen from the direct comparison shown in Fig..~\ref{fig:atmo_correlation}.

\begin{figure}
    \centering
    \includegraphics[width=\columnwidth]{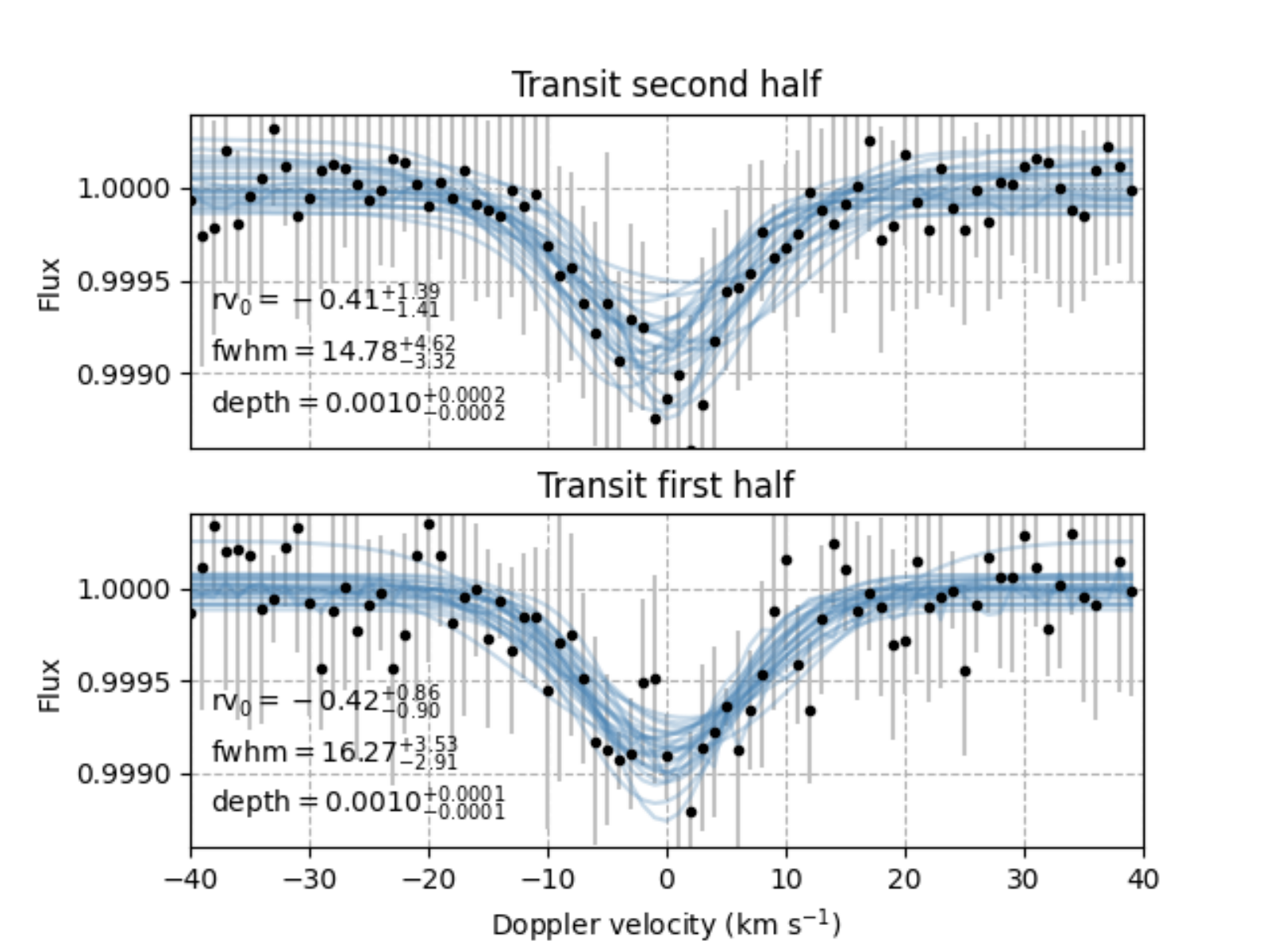}
    \caption{Averaged atmospheric signal in the first (phase [-0.02:0.0]) and second part (phase [0.0:0.02]) of the transit with relative fit using the unfiltered data.}
    \label{fig:atmo_two_halves_noSG}
\end{figure}

Also the overall variations during the transit (Fig.~\ref{fig:atmo_two_halves_noSG}), arising from comparing the atmospheric signal in the first and second half of the transit, are similar to those found in the filtered data (Fig.~\ref{fig:atmo_two_halves}). For example, in the unfiltered data the FWHM decreases from 16.27$^{+3.5}_{-2.9}$ to 14.78$^{+4.6}_{-3.3}$~\kms, while in the filtered data it decreases from 15.46$^{+1.4}_{-1.6}$ to 13.96$^{+0.9}_{-0.9}$~\kms: in both cases the decrease amounts to 1.5~\kms, while the slightly larger FWHM values found in the unfiltered dataset may be due to the excess noise.

We can reliably confirm that the Savitzky-Golay filter behaves as expected, preserving the signal's behaviour and lowering the uncertainties thanks to the cleaned out noise.
\end{appendix}

\end{document}